\documentclass[epj]{svjour}
\usepackage{graphicx}
\usepackage{dcolumn}
\usepackage{bm}
\usepackage{cite}
\usepackage{amsfonts}
\usepackage{amsmath}
\usepackage{amssymb}
\usepackage{bbold}
\usepackage{bbm}

\usepackage{array,multirow}

\usepackage[utf8]{inputenc}
\usepackage{upgreek}

\usepackage{tikz}
\usetikzlibrary{positioning,shapes}

\usepackage{hyperref}
\hypersetup{colorlinks=true,
citecolor=blue,
linkcolor=brown,
urlcolor=blue,
}

\def\be{\begin{equation}}
\def\ee{\end{equation}}
\def\ba{\begin{array}{c}}
\def\ea{\end{array}}
\def\p{\partial}
\def\ben{$$}
\def\een{$$}

\newcommand{\bea}{\begin{eqnarray}}
\newcommand{\eea}{\end{eqnarray}}

\newcommand{\bbr}{\br\!\br}
\newcommand{\kkt}{\kt\!\kt}
\newcommand{\pbr}{\prec\!}
\newcommand{\pkt}{\!\!\succ\,\,}
\newcommand{\kt}{\rangle}
\newcommand{\br}{\langle}

\begin{document} 
\title{Non-Hermitian coupled cluster method for non-stationary 
systems and its interaction-picture reinterpretation}
\titlerunning{Non-Hermitian CCM for non-stationary 
systems and its interaction-picture reinterpretation}
\author{R.~F.~Bishop\inst{1,2,3} \and M.~Znojil\inst{4,5,6}
}                     
%
%
\institute{Department of Physics and Astronomy, Schuster Building, 
The University of Manchester, Manchester, M13 9PL, United Kingdom \\
\email{raymond.bishop@manchester.ac.uk}
\and Quantum Systems Engineering Research Group, Department of Physics, 
Loughborough University, Leicestershire, LE11 3TU, United Kingdom
\and School of Physics and Astronomy, University of Minnesota, 
116 Church Street SE, Minneapolis, Minnesota 55455, USA
\and The Czech Academy of Sciences, Nuclear Physics
Institute, Hlavn\'{\i} 130, 250 68  \v{R}e\v{z}, Czech Republic \\
\email{znojil@ujf.cas.cz}
\and 
Department of Physics, Faculty of Science, University of Hradec
Kr\'{a}lov\'{e},
Rokitansk\'{e}ho 62, 500 03 Hradec Kr\'{a}lov\'{e},
 Czech Republic
\and Institute of System Science, Durban University of Technology,
P. O. Box 1334, Durban 4000, South Africa
}
\date{Received: date / Revised version: date}
%
\abstract{
The interaction picture in a non-Hermitian realization is discussed 
in detail and considered for its practical
use in many-body quantum physics.  
The resulting non-Hermitian interaction-picture (NHIP) description
of dynamics, in which both the wave functions
and operators belonging to physical observables cease 
to remain constant in
time, is a non-Hermitian generalization of the traditional Dirac picture 
of standard  quantum  mechanics, which itself is widely used
in quantum field theory calculations.
Particular attention is paid here to the variational 
(or, better, bivariational) and dynamical (i.e., non-stationary)
aspects that are characteristic of the coupled cluster method (CCM)
techniques that nowadays form one of the most versatile and most 
accurate of all available formulations of quantum many-body theory. In so doing
we expose and exploit multiple parallels between the NHIP and the
CCM in its time-dependent versions.
} 
\maketitle

\section{Introduction\label{introduction}}

In a wide variety of  branches of 
applied quantum mechanics
one often needs to know,
with a reasonable numerical precision, the value and evolution of the time-dependent
wave function $\psi(t)$. Not surprisingly,
the construction of this function
is relatively straightforward
only for a few
exceptional, not too
complicated
(and hence also, generically, 
not very realistic)
self-adjoint Hamiltonians
$\mathfrak{h}$ in the Schr\"{o}dinger equation
(in units where $\hbar=1$ throughout),
 \begin{equation}
 {\rm i}\partial_t \psi(t) = \mathfrak{h}\,\psi(t)\,.
 \label{jednicka}
 \end{equation}
The task usually requires the use of one of the available 
sophisticated  (e.g., typically, perturbative or variational)
numerical methods.
In the present paper we recall 
two of the apparently very different 
strategies of the variational class, with the express intention 
of showing that
behind the apparent formal differences one can also find
multiple parallels that can lead to new ideas and 
novel constructive approaches.

The present study has its roots 
in our earlier investigation \cite{Bishop-Znojil_2014} in which
we succeeded in comparing some alternative 
methods of solving
eq.~(\ref{jednicka}) 
in the stationary regime, 
in each of which the 
construction remains reducible 
to the diagonalization of the Hamiltonian
$\mathfrak{h}$.
In particular we revealed a close
structural parallelism between the powerful and versatile 
coupled cluster method (CCM) approach 
to the diagonalization of  
operator $\mathfrak{h}$ 
(and see, e.g., refs.~\cite{Arponen_1983,Arp-Bish-Paj_1987a,Arp-Bish-Paj_1987b,Bish-Arp-Paj_1989,Bishop_1991} for details 
of the formalism) 
and the successful interacting boson model (IBM) 
\cite{Arima-Iachello_1976,Arima-Iachello_1978,Arima-Iachello_1979}.  
The CCM
is, by now, well known and very widely used in 
such fields as atomic 
and molecular physics, nuclear physics, quantum chemistry 
(in which it forms the ``gold standard'' for accuracy), and 
many branches of condensed matter physics, while the IBM
offers an efficient tool for the calculation of the 
low-lying spectra of energies $E$
of the heaviest stable atomic nuclei 
\cite{Arima-Iachello_1976,Arima-Iachello_1978,Arima-Iachello_1979}.

In the present paper our aim is to demonstrate that 
the above-mentioned parallelism finds its further natural
extension to the domain of dynamics
in which the quantum systems in question are non-stationary.
We shall first recall the known 
CCM\,$\leftrightarrow$\,IBM parallels  \cite{Bishop-Znojil_2014},
and then show how to generalize them. We do so
along lines that are guided by the expectation 
of a mutual enrichment.
Explicitly, for the numerical constructions of energies $E$ 
and wave functions $\psi(t)$, 
we pay specific attention here to 
two specific alternative strategies, viz, the 
various versions of the CCM approach
(see, e.g., ref.~\cite{Bishop_1998} for 
a compact introductory review) 
and to standard (or ``textbook") quantum theory in its so called 
three-Hilbert-space formulation \cite{Znojil_SIGMA_2009,Bagarello-et-al-book_2015}.
For a comprehensive introduction to 
some of the most successful 
phenomenological applications of the three-Hilbert-space formalism
the reader is also referred to 
refs.~\cite{Bender-rpp_2007,Mostafazadeh-ijgmmp_2010}.

We anticipate that the future impact of 
our considerations might
range from   
a purely mathematical 
guarantee of the computational feasibility
of model-building, up to a deeply physical
reinterpretation of relationships between 
alternative formulations of many-body 
quantum mechanics.
We will emphasize
certain emergent features of 
connections between different means of 
description and between the 
simplifying assumptions and dynamics-simulating techniques.

\section{The fundamentals of the CCM approach \label{CCM-fundamentals}}


{\em None\,} of the existing CCM-based calculations leaves 
the traditional theoretical framework of quantum mechanics. The correspondence between the IBM and CCM formulations
of quantum mechanics may be one-to-one. 
Indeed, both formulations employ the factorized Ansatz
 \be
 \fontdimen16\textfont2=3.1pt
 \fontdimen17\textfont2=3.1pt
 \psi = \Omega\,\psi_0\,,
  \label{fakto}
  \ee
in which the auxiliary (and, generally, non-unitary) operator $\Omega$
is intended to serve as the source of the quantum correlations which 
are unaccounted for in the original approximation $\psi_0$. One of the most natural mathematical 
requirements imposed upon 
the operator $\Omega$
is that it should be {\em invertible}.
Thus, the element $\psi$ of an initial
physical Hilbert space
${\cal
H}^{(\rm initial)}$ is treated as an image of a, presumably,
{\em perceivably simpler\,} reference state
$\psi_0$, which may itself belong, in principle, to another, more
user-friendly, auxiliary Hilbert space 
${\cal H}^{\rm (user\!-\!friendly)}$.
The original Hamiltonian $\mathfrak{h}$ of 
eq.~(\ref{jednicka}) belonging to 
${\cal H}^{(\rm initial)}$ is similarly replaced
by its isospectral (and, in general, non-Hermitian) avatar,
 \be
 H=\Omega^{-1}\mathfrak{h}\Omega\,,
 \label{obshamst}
 \ee
that belongs to the auxiliary Hilbert space 
${\cal H}^{\rm (user\!-\!friendly)}$.

In the CCM setting, the auxiliary operator $\Omega$ 
in the Ansatz of eq.~(\ref{fakto}) takes
the very specific exponentiated form, 
 \be 
 \Omega=\exp S\,.
 \label{expS}
 \ee 
This characteristic features of the method henceforth 
ensures that the Goldstone linked cluster
theorem \cite{Goldstone_1957}
is automatically satisfied for the system under study 
at any level of 
approximation for the operator $S$, as we explain
in more detail below in sec.~\ref{CCM-fundamentals}.  In turn, 
this guarantees the very important property for all
many-body systems that at all such levels
the system is both size-extensive and size-consistent 
\cite{Bishop_1998}, 
where size-extensivity is the property that the leading term in the energy
of an $N$-particle self-bound system scales linearly 
with $N$ as $N \to \infty$, and 
where size-consistency implies that a many-body wave function dissociates 
correctly into non-interacting
fragments under infinite separation of the fragments.

The original version of the CCM, as invented
independently by Coester and K\"{u}mmel 
\cite{Coester_1958,Coester-Kummel_1960} 
and \v{C}\'{i}\v{z}ek \cite{Cizek_1966,Cizek_1969}, is nowadays
referred to as the normal CCM (NCCM), in order to distinguish it
from the later extended (ECCM) version introduced by Arponen
\cite{Arponen_1983}.  For present purposes most of what we discuss
here is equally relevant to both versions, although we mostly have 
the NCCM in mind for specific applications.

\subsection{The normal coupled cluster method (NCCM)} \label{NCCM-fund}

The exact ket and bra wave functions for an arbitrary quantum many-body
system are defined in ${\cal H}^{(\rm initial)}$ to be $|\psi(t)\rangle$ and
\begin{equation}
 \langle\tilde{\psi}(t)|
 \equiv \frac{\langle\psi(t)|}{\langle\psi(t)|\psi(t)\rangle}\,,
 \label{bra-norm}
\end{equation}
respectively, at an arbitrary time $t$.  We assume that the Hilbert 
space ${\cal H}^{(\rm initial)}$ for the
system may be described in terms of a normalized,
stationary reference (or model) state $|\psi_0\rangle$, 
(i.e., with $\langle\psi_0|\psi_0\rangle=1$), 
just as in eq.~(\ref{fakto}). Within the CCM
we further assume that the reference state acts as a cyclic vector
for a corresponding set of {\em mutually commuting\,} multi-configurational
creation operators $\{{C}_I^+\}$,
\begin{equation}
 [C_I^+,C_J^+]=0\,;\quad \forall I,J\,,  
 \label{creators-commute}
\end{equation}
such that $|\psi_0\rangle$ acts as a {\em generalized vacuum state\,} 
with respect to them,
\begin{equation}
 C_I^-|\psi_0\rangle = 0 = \langle\psi_0|C_I^+ \,;\quad\forall I\neq 0\,,
 \label{gen-vac}
\end{equation}
in a notation where we define $C_I^-\equiv (C_I^+)^\dag$ and
$C_0^+\equiv\mathbbm{1}$, 
the unit vector.
The reference state $|\psi_0\rangle$ must also be chosen to be non-orthogonal 
to the actual wave function $|\psi(t)\rangle$ of the system,
\begin{equation}
 \langle\psi_0|\psi(t)\rangle \neq 0 
 \,;\quad \forall t\,.
 \label{state-norm}
\end{equation}

The index $I$ is a set index and, in general, the multi-configurational 
creation operator $C_I^+$ comprises
a product of single-particle operators, as we illustrate below 
in sec.~\ref{Raa-sub} for a specific example.
The set $\{I\}$ is complete in the usual sense
that the set of states $\{C_I^+|\psi_0\rangle\}$ provides a complete basis
for the ket Hilbert space. It is also
convenient to choose the basis to be orthonormalized,
such that  
\begin{equation}
 \langle\psi_0|C_I^-C_J^+|\psi_0\rangle=\delta_{IJ} \,,
 \label{creator-norm}
\end{equation}
where $\delta_{IJ}$ is an appropriately
defined Kronecker symbol.  In this case we have that
the identity operator in the Hilbert space may be resolved as
\begin{equation}
 \sum_I C_I^+|\psi_0\rangle\langle\psi_0| C_I^- = \mathbbm{1}  
 =|\psi_0\rangle\langle\psi_0| + \sum_{I\neq0} C_I^+|\psi_0\rangle\langle\psi_0| C_I^-\,.
 \label{identity-resolution}
\end{equation}
The ket and bra states of the many-body system are now formally parametrized 
{\em independently\,} in the NCCM as follows,
\begin{equation}
 |\psi(t)\rangle \equiv {\rm e}^{k(t)}{\rm e}^{S(t)}|\psi_0\rangle \,,\quad
 \langle\tilde{\psi}(t)| \equiv {\rm e}^{-k(t)}\langle\psi_0|
 \tilde{S}(t){\rm e}^{-S(t)}\,,
 \label{NCCM-states}
\end{equation}
in terms of the $c$-number function $k(t)$, and the 
NCCM time-dependent correlation operators $S(t)$ and
$\tilde{S}(t)$, which are themselves now decomposed as follows,
\begin{equation}
 S(t) = \sum_{I\neq0}s_I(t)C_I^+\,,\quad
 \tilde{S}(t)=\mathbbm{1}+\sum_{I\neq0}\tilde{s}_I(t)C_I^-\,.
 \label{NCCM-ops}   
\end{equation}
The scale factor $k(t)$ is needed for the 
temporally evolving intermediate normalization
(i.e., $\langle\psi_0|\psi(t)\rangle={\rm e}^{k(t)}$) 
but is irrelevant henceforth 
since it cancels from any expectation value
[and the normalization condition 
$\langle\tilde{\psi}(t)|\psi(t)\rangle =1$ 
from eq.~(\ref{bra-norm})
is now preserved for all times $t$] by the specific
parametrizations of eqs.~(\ref{NCCM-states}) and (\ref{NCCM-ops}).
Since $S(t)$ and $\tilde{S}(t)$ are
henceforth treated as independent operators, the Hermiticity
relation between corresponding ket and bra states may be violated
when subsequent approximations are made, typically by
truncating the complete set of configurations $\{I\}$ in the
sums of eq.~(\ref{NCCM-ops}). This
shortcoming is far outweighed by
the fact that the CCM parametrizations of 
eqs.~(\ref{NCCM-states}) and (\ref{NCCM-ops}) 
always {\em exactly\,} satisfy the very important Hellmann-Feynman 
theorem \cite{Hellmann_1935,Feynman_1939} at 
all such levels of approximation \cite{Bishop_1998}.
It is precisely this feature that guarantees the robustness
and accuracy of numerical results obtained within the
CCM framework.

The expectation value $\overline{Q}(t)$ of an arbitrary
physical observable described by the Hermitian operator 
$\mathfrak{q}(t)$ in ${\cal H}^{(\rm initial)}$,
which may itself have an intrinsic time dependence,
may thus be expressed within the NCCM wholly in terms
of the set of $c$-number correlation coefficients
$\{s_I(t),\tilde{s}_I(t)\}$ as follows,
\begin{equation}
 \br\mathfrak{q}(t)\kt=\overline{Q}(t) = \overline{Q}[s_I,\tilde{s}_I;t] 
 \equiv  \langle\psi_0|\tilde{S}(t){\rm e}^{-S(t)}\mathfrak{q}(t){\rm e}^{S(t)}
 |\psi_0\rangle\,.
 \label{NCCM-ev:a}
\end{equation}
Such an expectation functional
is evaluated by employing 
the nested commutation expansion for the
similarity transform $Q(t)$
[c.f., eq.~(\ref{obshamst})] which is just the NCCM avatar in ${\cal H}^{\rm (user\!-\!friendly)}$ 
of the operator $\mathfrak{q}(t)$ in ${\cal H}^{(\rm initial)}$,
\begin{subequations}
 \begin{align}
  Q(t) &\equiv {\rm e}^{-S(t)}\mathfrak{q}(t){\rm e}^{S(t)}
  \label{CCM-gen-sim-Xm}\\
  &= \sum_{n=0}^{\infty}
  \frac{1}{n!}\, [\mathfrak{q}(t),S(t)]_{n}\,.
  \label{nestsum:a}
 \end{align}
\end{subequations}
Here, the $n$-fold nested commutators $[\mathfrak{q},S]_{n}$ are defined iteratively,
\be
 [\mathfrak{q},S]_{n} = [[\mathfrak{q},S]_{n-1},S]\,, \quad 
 [\mathfrak{q},S]_{0} \equiv \mathfrak{q}\,.
 \label{nestsum:b}
\ee
The similarity-transformed Hamiltonian,
$H(t)\equiv{\rm e}^{-S(t)}\mathfrak{h}{\rm e}^{S(t)}$,
lies at the heart of the NCCM.
Its parametrizations of eqs.~(\ref{NCCM-states}) and (\ref{NCCM-ops}) are
specifically chosen so that the otherwise infinite sum in
eq.~(\ref{nestsum:a}) will actually terminate exactly at a low finite order
for any operator $\mathfrak{q}$ all of whose terms involve only
a product of a finite number of single-particle operators,
such as is almost always the case 
for the operators of interest. The reason
for this termination lies simply in the twin facts that all
components of the operator $S(t)$ in the expansion of eq.~(\ref{NCCM-ops}) commute
with one another from eq.~(\ref{creators-commute}), 
and the basic single-particle operators
form a Lie algebra, together with the model state $|\psi_0\rangle$ being
a vacuum state for the operator set $\{C_I^+\}$, 
as in eq.~(\ref{gen-vac}) (and see, e.g.,
refs.~\cite{Bishop_1991,Bishop_1998} for further details). 

The above same key
feature of the NCCM is also responsible for all terms in the
expansion of $\overline{Q}(t)$ from eq.~(\ref{NCCM-ev:a})
for any physical operator $\mathfrak{q}$ being {\em linked\,}.
Thus, we may write the NCCM expectation value functional
$\overline{Q}(t)$ in the schematic form,
\begin{equation}
  \overline{Q}[s_I,\tilde{s}_I;t]=\sum_{n=0}^{\infty}\frac{1}{n!}\sum_J
  \sum_{K_1\neq0}\cdots\!\sum_{K_n\neq0}
  \br J|\mathfrak{q}(t)|K_1,\cdots\! ,K_n\kt_{\mbox{\scriptsize{\rm NCCM}}}
  \tilde{s}_{J}s_{K_n}\cdots s_{K_1}\,
  \label{NCCM-ev:b}
\end{equation}
where, by comparison with eqs.~(\ref{nestsum:a}) and (\ref{nestsum:b}),
the NCCM matrix elements may be written as
\begin{equation}
  \br J|\mathfrak{q}(t)|K_1,\cdots\! ,K_n\kt_{\mbox{\scriptsize{\rm NCCM}}}
  \equiv \br\psi_0|C_J^-\,\mathfrak{q}(t)\,C_{K_1}^+\cdots C_{K_n}^+ 
  |\psi_0\kt_{\mbox{\scriptsize{$\mathcal{L}$}}}\,.
  \label{NCCM-ME_L}
\end{equation}
The suffix $\mathcal{L}$ in eq.~(\ref{NCCM-ME_L}) 
indicates the NCCM {\em linked\,} structure implied
by the nested commutator sum, viz., that from each group of particles
characterised by each of the configuration-space indices $K_j$, the 
amplitude $s_{K_j}$ associated with the cluster creation operator 
$C_{K_j}^+$, must have at least one particle line connected
to the operator $\mathfrak{q}$.  After expanding 
eq.~(\ref{NCCM-ev:a}) in powers of $S$ by the use of eq.~(\ref{nestsum:a}), 
only such terms are retained where 
at least one link (or contraction, due to a non-vanishing commutator term) 
exists between each $S$ operator and the 
operator $\mathfrak{q}$. The sum over $n$
in eq.~(\ref{NCCM-ev:b}) will extend again to a finite limit depending
on the operator $\mathfrak{q}$.

It is precisely this linkedness feature
(specifically for the energy expectation value functional
$\overline{H}(t)$), as expressed in 
eqs.~(\ref{NCCM-ev:b}) and (\ref{NCCM-ME_L}),
that implies immediately that the NCCM automatically
satisfies exactly the Goldstone linked-cluster
theorem \cite{Goldstone_1957} at any level of truncation in the expansions of
eq.~(\ref{NCCM-ops}) for the correlation operators. 
In turn this also immediately
guarantees the size-extensivity of the method
at all such levels of approximation. Thus, in the solution
of the explicit evolution equations that we will now derive
for the basic NCCM amplitudes $\{s_I(t),\tilde{s}_(t)\}$, 
the only approximation
that is ever made in any CCM calculation is to decide
which configurations $\{I\}$ are to be retained in the expansions
of eq.~(\ref{NCCM-ops}).

The NCCM coefficients $\{s_I(t),\tilde{s}_I(t)\}$ 
are obtained 
by introducing the action functional $\mathcal{A}$, 
\be
 \mathcal{A}\equiv\int_{t_0}^{t_1} {\rm d}t\,
 \langle\tilde{\psi}(t)|\left( {\rm i}\partial/\partial t 
 -\mathfrak{h}\right)|\psi(t)\,.
 \label{action-def}
\ee 
The (bivariational) stationarity principle,
\be
 \frac{\partial\mathcal{A}}{\partial\langle\tilde{\psi}(t)|}=0
 =\frac{\partial\mathcal{A}}{\partial|{\psi}(t)\rangle}\,,
 \label{action-min-gen}
\ee 
with respect to all independent variations in the bra and ket states,
subject to the conditions $\delta\langle\tilde{\psi}(t_i)|=0=
\delta|\psi(t_i)\rangle$ where $i=0,1$, is seen to
be equivalent to the time-dependent Schr\"{o}dinger equations,
\be
 {\rm i}\partial_t |\psi(t)\rangle = \mathfrak{h}\,\psi(t)\,,\quad
 -{\rm i}\partial_t\langle\tilde{\psi}(t)|=\langle\tilde{\psi}(t)|\mathfrak{h}\,,
 \label{TDSEs}
\ee
analogous to eq.~(\ref{jednicka}). Insertion of the NCCM 
parametrizations of eqs. (\ref{NCCM-states}) and (\ref{NCCM-ops}) into
eq.~(\ref{action-def}) yields the result,
\be
 \mathcal{A}=\int_{t_0}^{t_1} {\rm d}t\,\bigg\{
 {\rm i}\sum_{I\neq0}\Big(\tilde{s}_I\frac{{\rm d}s_I}{{\rm d}t}\Big)
 -\overline{H}[s_I,\tilde{s}_I]\bigg\}
 =\int_{t_0}^{t_1} {\rm d}t\,\bigg\{-
 {\rm i}\sum_{I\neq0}\Big(\frac{{\rm d}\tilde{s}_I}{{\rm d}t}s_I\Big)
 -\overline{H}[s_I,\tilde{s}_I]\bigg\}\,.
 \label{NCCM-action}
\ee
The independent (bivariational) stationarity conditions,
$\partial\mathcal{A}/\partial\tilde{s}_I=0=\partial\mathcal{A}/\partial s_I$,
then yield the  evolution equations for the 
NCCM cluster correlation coefficients,
\be
 {\rm i}\frac{{\rm d}s_I}{{\rm d}t}
 =\frac{\partial \overline{H}}{\partial \tilde{s}_I}\,,\quad
 -{\rm i}\frac{{\rm d}\tilde{s}_I}{{\rm d}t}
 =\frac{\partial \overline{H}}{\partial s_I}\,;
 \quad \forall I\neq 0\,,
 \label{NNCCM-evol-eqs}
\ee
in terms of the energy expectation functional 
$\overline{H}$, given
as in eq.~(\ref{NCCM-ev:a}) with the replacement
$\mathfrak{q}(t)\to\mathfrak{h}$. We recognise these equations as 
(being closely related to)
the {\em classical\,} Hamilton’s equations of motion for the 
NCCM cluster correlation coefficients.  Thus, the coefficients
$s_I$ and $\tilde{s}_I$ form {\em canonically conjugate pairs\,} in 
the usual sense of classical mechanics.

We can now also  easily derive the equation of motion for the expectation value
${\overline Q}(t)\equiv {\overline Q}[s_{I},\tilde{s}_{I};t]$ of an 
arbitrary operator $\mathfrak{q}(t)$ in 
${\cal H}^{(\rm initial)}$ given by eq.~(\ref{NCCM-ev:a}).
Using eq.~(\ref{NNCCM-evol-eqs}), 
we rewrite the usual chain-rule relation,
 \be
 \frac{{\rm d}{\overline Q}}{{\rm d}t}=\frac{\partial{\overline Q}}{\partial t}+
 \sum_{I\neq0}\left(\frac{\partial{\overline Q}}{\partial s_I}\,
 \frac{{\rm d}{s_I}}{{\rm d}t}+\frac{\partial{\overline Q}}{\partial \tilde{s}_I}\,
 \frac{{\rm d}{\tilde{s}_I}}{{\rm d}t}
 \right)\,,
 \ee
in the suggestive form,
 \be
 \frac{{\rm d}{\overline Q}}{{\rm d}t}=\frac{\partial{\overline Q}}{\partial t}+
 \{{\overline Q},{\overline H}\}\,,
 \label{NCCM_EOM-for-Vbar}
 \ee
in terms of a generalized {\em classical Poisson bracket\,}, defined between
the expectation values $\overline A$ and $\overline B$ of two arbitrary operators
in ${\cal H}^{\rm (user\!-\!friendly)}$ as
follows,
 \be
 \{{\overline A},{\overline B}\}\equiv \frac{1}{\rm i} \sum_{I\neq0}\left(
 \frac{\partial{\overline A}}{\partial s_I}\,\frac{\partial{\overline B}}{\partial \tilde{s}_I}-
 \frac{\partial{\overline A}}{\partial \tilde{s}_I}\,\frac{\partial{\overline B}}{\partial s_I}
 \right)\,.
 \label{NCCM_Poisson-a}
 \ee
Equation (\ref{NCCM_EOM-for-Vbar}) is just the well-known classical
equation of motion in the canonical formalism.
It is clearly the classical, {\em exactly mapped\,}, counterpart
of the usual quantum-mechanical Heisenberg equation of motion for
the operator $\mathfrak{q}(t)$ in the original Hilbert
space ${\cal H}^{(\rm initial)}$,
 \be
 \frac{{\rm d}\mathfrak{q}}{{\rm d}t}=\frac{\partial\mathfrak{q}}{\partial t}+
 \frac{1}{\rm i}[\mathfrak{q},\mathfrak{h}]\,,
 \label{Heisenber-EOM}
 \ee
with the lower-case operator $\mathfrak{q}(t)$ and its upper-case NCCM
similarity-transformed
avatar operator $Q(t)$ related  by eq.~(\ref{CCM-gen-sim-Xm})

The above results can be made even more suggestive 
by choosing new basic variables to be
the generalized $c$-number fields $\phi_I$ and their canonically conjugate 
$c$-number generalized momentum densities $\pi_I$, which are defined
as follows,
\begin{subequations}
 \be
 \phi_I\equiv 2^{-1/2}(s_I+\tilde{s}_I)\,;\quad
 \pi_I\equiv -{\rm i}\,2^{-1/2}(s_I-\tilde{s}_I)\,.
 \label{field-momentum-Xn}
 \ee
The inverse transformations are thus simply given by
 \be
 s_I\equiv 2^{-1/2}(\phi_I+{\rm i}\pi_I)\,;\quad
 \tilde{s}_I\equiv 2^{-1/2}(\phi_I-{\rm i}\pi_I)\,.
 \label{inv-field-momentum-Xn}
 \ee
\end{subequations}
It is now easy to show that Eq.~(\ref{NCCM_Poisson-a}) in the new variables
takes the following form,
 \be
 \{{\overline A},{\overline B}\}= \sum_{I\neq0}\left(
 \frac{\partial{\overline A}}{\partial \phi_I}\,\frac{\partial{\overline B}}{\partial \pi_I}-
 \frac{\partial{\overline A}}{\partial \pi_I}\,\frac{\partial{\overline B}}{\partial \phi_I}
 \right)\,.
 \label{Poisson-b}
 \ee
The equations of motion (\ref{NNCCM-evol-eqs}) 
lead to their equivalent
counterparts for the new variables,
 \begin{subequations}
 \begin{align}
 \frac{\rm d\phi_I}{{\rm d}t} &=\frac{\partial{\overline H}}{\partial\pi_I}
 =\{\phi_I,{\overline H}\}\,;
 \quad \forall I \neq 0\,,
 \label{EOM-for-pi_I}\\
 \frac{\rm d\pi_I}{{\rm d}t} &=-\frac{\partial{\overline H}}{\partial \phi_I}
 =\{\pi_I,{\overline H}\}\,;
 \quad \forall I \neq 0\,.
 \label{EOM-for-phi_I}
 \end{align}
 \end{subequations}
The (completely classical) phase space $\{\phi_I,\pi_I\}$ has the 
canonical symplectic structure,
 \be
 \begin{split}
     \{\phi_I,\pi_J\} &=\delta_{I,J}\,;\quad \forall I\neq0
     \,, J\neq0\,,\\
     \{\phi_I,\phi_J\} &=0=\{\pi_I,\pi_J\}\,.
 \end{split}
 \label{symplectic}
 \ee

In order to complete this description of an exact
{\em classicization\,} of an arbitrary quantum many-body theory,
we note that the expectation value
of the commutator between an {\em arbitrary\,} pair of operators
$\mathfrak{a}$ and $\mathfrak{b}$ in the original Hilbert
space ${\cal H}^{(\rm initial)}$ is also {\em exactly\,}
mapped into its corresponding Poisson bracket of eq.~(\ref{NCCM_Poisson-a})
or eq.~(\ref{Poisson-b}). We may
use eq.~(\ref{NCCM-ev:a}) to write the 
expectation value of the product $\mathfrak{ab}$ of the two operators
as
 \be
 \br\mathfrak{ab}\kt=
 \langle\psi_0|\tilde{S}(t){\rm e}^{-S(t)}\mathfrak{ab}{\rm e}^{S(t)}
 |\psi_0\rangle =
 \langle\psi_0|\tilde{S}(t){\rm e}^{-S(t)}\mathfrak{a}{\rm e}^{S(t)}
 {\rm e}^{-S(t)}\mathfrak{b}{\rm e}^{S(t)}
 |\psi_0\rangle\,,
 \ee
and hence, using eq.~(\ref{CCM-gen-sim-Xm}), as
 \be
 \br\mathfrak{ab}\kt=
 \br\psi_0|\tilde{S}(t)A(t)B(t)|\psi_0\kt=
 \overline{AB}\,.
 \ee
By inserting a resolution of the identity of the form of 
eq.~(\ref{identity-resolution}) between the operators
$A(t)$ and $B(t)$ in the middle 
expression in the above equation, we find
\begin{align}
\begin{split}
 \overline{AB} &=
 \sum_{I}\br\psi_0|\tilde{S}(t)A(t)C_I^+|\psi_0\kt
 \br\psi_0|C_{I}^{-}B(t)|\psi\kt\\
 &= \sum_{I}\br\psi_0|\tilde{S}(t)[A(t),C_I^{+}]|\psi\kt
 \br\psi_0|C_{I}^{-}B(t)|\psi_0\kt+
 \sum_{I}\br\psi_0|\tilde{S}(t)C_I^{+}A(t)|\psi_0\kt
 \br\psi_0|C_{I}^{-}B(t)|\psi_0\kt
 \,.
\end{split} 
\label{inter}
\end{align}
The term with $I=0$ in the sum in the first term of
the above equation is identically zero, since $C_0^+\equiv \mathbbm{1}$.
Next, if we use the explicit form for the expectation value
of an arbitrary operator given by eq.~(\ref{NCCM-ev:a}), together
with the NCCM forms for the operators $S(t)$ and $\tilde{S}(t)$ 
given by eq.~(\ref{NCCM-ops}),
we finally derive the following relations 
for the remaining terms,
\begin{subequations}
  \begin{align}
  \frac{\partial{\overline Q}}{\partial\tilde{s}_I}&=\br\psi_0|C_{I}^{-}Q(t)|\psi_0\kt\,;
  \quad\forall I\neq 0\,,
  \label{partial-of-Qbar-wrt-stilde_I}\\
  \frac{\partial{\overline Q}}{\partial s_I}&=\br\psi_0|\tilde{S}(t)
  [Q(t),C_{I}^{+}]|\psi_0\kt\,;
  \quad\forall I\neq 0\,.
  \label{partial-of-Qbar-wrt-s_I}
  \end{align}
\end{subequations} 
Substitution of eqs.~(\ref{partial-of-Qbar-wrt-stilde_I}) 
and (\ref{partial-of-Qbar-wrt-s_I}) into 
the first term of the second line of eq.~(\ref{inter})
then yields the following result,
 \be
 \br\mathfrak{ab}\kt=
 \overline{AB}=\sum_{I\neq0}
 \frac{\partial{\overline A}}{\partial s_I}\,\frac{\partial{\overline B}}{\partial \tilde{s}_I}+
 \sum_{I}\sum_J\br\psi_0|\tilde{S}(t)C_I^{+}C_J^{+}|\psi_0\kt
 \br\psi_0|C_J^{-}A(t)|\psi_0\kt\br\psi_0|C_{I}^{-}B(t)|\psi_0\kt\,
 \label{exp-value_for-product-AB}
 \ee
where in the second term of 
the second line of eq.~(\ref{inter}) we have inserted 
a resolution of the identity operator of the form of
eq.~(\ref{identity-resolution}) between the operators $C_I^+$ and
$A(t)$. The last term in eq.~(\ref{exp-value_for-product-AB}) is
symmetric under the interchange $A\rightleftarrows B$, as 
$[C_I^+,C_J^+] =0$ from eq.~(\ref{creators-commute}), 
and hence we arrive at
the desired relation,
 \be
 \br[\mathfrak{a},\mathfrak{b}]\kt=
 \overline{[A,B]}=
 \overline{AB}-\overline{BA}=\sum_{I\neq0}\left(
\frac{\partial{\overline A}}{\partial s_I}\,\frac{\partial{\overline B}}{\partial \tilde{s}_I}-
 \frac{\partial{\overline A}}{\partial \tilde{s}_I}\,\frac{\partial{\overline B}}{\partial s_I}
 \right)={\rm i}\{{\overline A},{\overline B}\}\,.
 \label{NCCM_commutator-map-to-Poisson}
 \ee
It is very important to note that the consistency of the
classical and quantum formulations induced by the NCCM
can thus be understood as a manifestation
of the correspondence principle (or Ehrenfest theorem)
in a suitably generalized form. 

In partial summary, we have thus been led to a key result of the NCCM
formalism, namely that our (rather arbitrarily specified) quantum many-body
problem in a Hilbert space has been exactly mapped onto 
the classical Hamiltonian mechanics 
for the complex $c$-number amplitudes $\{s_I,\tilde{s}_I\}$, or equivalently
the canonical fields $\{\phi_I\}$ and their conjugate momentum densities
$\{\pi_I\}$, in a classical NCCM phase space. These fields are 
themselves defined as many-body amplitudes in the configuration space 
labelled by the indices $\{I\}$ that describe
the particular subsets (or clusters) of particles under consideration. 
In the modern terminology of classical mechanics this phase space is a 
symplectic differentiable manifold \cite{Abraham-Marsden-book_1978}.
The differentiability is just a consequence of the fact that the $c$-number
parameters $\{s_I,\tilde{s}_I\}$ or, equivalently, $\{\phi_I,\pi_I\}$, are 
continuous complex numbers, with respect to which one is allowed to
take (possibly functional) derivatives.  The symplectic nature of the phase space
is a direct consequence of the existence of the generalized Poisson bracket, which
is just a skew-symmetric bilinear form that can be used to define a Hamiltonian vector
field in the tangent space of the manifold \cite{Abraham-Marsden-book_1978}.
The set of trajectories defined by the equations of motion will
fill the whole of the dynamically allowed region of the phase space.

The exact classicization mapping opens up
the possibility of being able to exploit or to extend all of the techniques developed
in classical mechanics for use in the quantum many-body problem. For example, one 
can make easy contact with {\em conservation laws} and the 
associated sum rules by using the 
Noether currents. It is also intuitively apparent from the existence of this mapping onto
classical mechanics that the $c$-number NCCM amplitudes $\{\phi_I,\pi_I\}$  or 
$\{s_I,\tilde{s}_I\}$, which
completely characterise and decompose our many-body problem, may be viewed as a set
of {\em generalized mean fields\,} which describe each 
subsystem of particles in the interacting many-body system as labelled 
by the configuration-space indices $\{I\}$.

Let us now return briefly to the non-Hermitian nature of the NCCM approach. 
The requirement that the physical bra and ket states in the original Hilbert space
${\cal H}^{(\rm initial)}$ should be Hermitian conjugates of one another,
implies from eqs.~(\ref{bra-norm}) and (\ref{NCCM-states}) that
 \be
 \br\psi_0|\tilde{S}
 =\frac{\br\psi_0|{\rm e}^{S^\dagger}{\rm e}^{S}}
 {\br\psi_0|{\rm e}^{S^\dagger}{\rm e}^{S}|\psi_0\kt}\,,
 \label{hermiticity}
 \ee
and hence to constraints between the set of amplitudes
$\{s_I,\tilde{s}_I\}$.  Thus, the observable physical space
is just a submanifold in the full NCCM symplectic phase space.
It is, however, an invariant submanifold in the sense that it comprises
entire trajectories, such that either none or all of the points of a 
trajectory belong to the physical submanifold, under the sole assumption that
the original Hamiltonian $\mathfrak{h}$ in ${\cal H}^{(\rm initial)}$ is 
Hermitian.  Temporal evolution thus leaves this submanifold invariant.
Even within this submanifold the respective pairs of amplitudes
$s_I$ and $\tilde{s}_I$ are not in general complex conjugates of one
another since the underlying NCCM similarity transformation is not unitary. These pairs of amplitudes are in no simple relationship to one another.
Correspondingly, neither the generalized fields $\phi_I$ nor their
canonically conjugate generalized momentaum densities $\pi_I$, as defined
in Eq.~(\ref{field-momentum-Xn}) will in general be real. It is only
by performing a suitable complex canonical transformation (i.e., a
symplectomorphism) that one can regain a description in terms of manifestly
real coordinates $\{\phi_{I}^{'},\pi_{I}^{'}\}$ instead of complex ones.

There are actually (at least) two different approaches to achieving
the above aim.  The first, and perhaps most direct, has been performed by 
Arponen \cite{Arponen_1997}, who first introduces the set of complex conjugate
NCCM amplitudes $\{s_I^*,\tilde{s}_I^*\}$, whereby the NCCM phase space is 
enlarged into a genuine complex manifold, but self-evidently now of too
large a dimensionality.   Thus, finally, the extra degrees of freedom 
are explicitly eliminated by using the Dirac bracket method and deriving
the induced symplectic structure, in order to construct the physical
manifold (otherwise known as the constraint surface or physical shell), in which the
remaining independent amplitudes are a minimal set of complex conjugate pairs or,
equivalently, a manifestly real set $\{\phi_{I}^{'},\pi_{I}^{'}\}$.  He shows
that the NCCM physical shell is actually a K\"ahler manifold, and the constraint 
functions themselves are of second class
\cite{Henneaux-Teitelboim-book_1994}, as is precisely the case in a gauge theory
where the constraint corresponds precisely to a complete gauge fixing, even though
in the NCCM case the constraints do not seem to correspond to any internal
(hidden) gauge symmetries of the problem. The resulting reduced phase space of the
physical shell is shown to be itself a complex manifold with a symplectic
structure, just like the original extended phase space. The symplectic structure 
is given by an induced symplectic two-form, whose general form is derived without
making any approximations or restricting assumptions. It has also been shown 
in ref.~\cite{Arponen_1997} that the NCCM star product is well
defined in the reduced phase space, or the constraint surface, where
the star product is defined to give the expectation of a product of operators
in terms of the expectations of individual operators, and where it is also
shown that it is implemented by a certain tensor field of type $(2,0)$. 

The second approach to ``re-Hermitizing'' the NCCM formulation by a canonical
transformation (i.e., a symplectomorphism) of the set of amplitudes 
$\{s_I,\tilde{s}_I\}$ (or, equivalently, $\{\phi_{I},\pi_{I}\}$)
has been performed more indirectly by diagonalizing the effective
Hamiltonian that governs small oscillations around the equilibrium ground state
\cite{Arp-Bish-Paj_1987b}. It has been explicity shown in 
ref.~\cite{Arp-Bish-Paj_1987b} that this diagonalization procedure
is equivalent to performing a canonical coordinate transformation 
into normal coordinates in the symplectic phase space. The normal
frequencies are also then precisely the excitation energies.  In terms 
of the new normal coordinates the Hamiltonian functional ${\overline H}$ 
thus recovers a manifestly Hermitian form that was lost in the original
NCCM form of the functional $\overline{H}[s_{I},\tilde{s}_{I}]$ due
to the nonunitary nature of the similarity transformation that 
lies at the heart of the NCCM.

\subsection{The extended coupled cluster method (ECCM)} \label{ECCM-fund}

Let us now briefly conclude our discussion of the CCM
fundamentals by considering the extension of the above ideas to
the ECCM. As we have seen, any NCCM expectation value functional
$\overline{Q}(t) = \overline{Q}[s_I,\tilde{s}_I;t]$ comprises 
only linked (or connected) terms.  The exponentiated form of the NCCM ket-state
correlation operator in eq.~(\ref{NCCM-states}) also guarantees
that all of the coefficients $\{s_I;\,\forall I\neq0\}$ are 
themselves fully linked.  However, the  corresponding canonically 
conjugate NCCM bra-state coefficients $\{\tilde{s}_I;\,\forall I\neq0\}$
{\em do\,} contain unlinked terms.  If and when needed, this can be  
remedied by  introducing the following ECCM
exponentiated form for the NCCM bra-state correlation operator,
$\tilde{S}(t)$,
\begin{equation}
  \tilde{S}(t) \equiv {\rm e}^{\tilde{\Sigma}(t)}\,;\quad
 \tilde{\Sigma}(t)=\sum_{I\neq0}\tilde{\sigma}_I(t)C_I^-\,. 
 \label{ECCM-Sigma-tilde-op}
\end{equation}
It is clear from eqs.~(\ref{creators-commute}), 
(\ref{NCCM-ops}) and (\ref{NCCM-ev:a}) 
that the NCCM cluster coefficients $\tilde{s}_I$ may be
expressed as $\tilde{s}_I=\br C_I^+\kt$, which in general
are {\em not\,} linked quantities.  
The linked pieces of these expectation values that
remain after the disconnected terms have been discarded
are precisely the new ECCM cluster coefficients,
$\tilde{\sigma}_I=\br C_I^+\kt_{\mbox{\scriptsize{\rm linked}}}$.

From eqs.~(\ref{gen-vac}) and (\ref{ECCM-Sigma-tilde-op}) we
deduce that $\tilde{\Sigma}(t)|\psi_0\kt=0$ so that we
may write the ECCM parametrizations of the ket and bra many-body
wave functions as follows,
\begin{equation}
 |\psi(t)\rangle \equiv {\rm e}^{k(t)}{\rm e}^{S(t)}|\psi_0\rangle
 = {\rm e}^{k(t)}{\rm e}^{S(t)}{\rm e}^{-\tilde{\Sigma}t)}
 |\psi_0\rangle \,,\quad
 \langle\tilde{\psi}(t)| \equiv {\rm e}^{-k(t)}\langle\psi_0|
 {\rm e}^{\tilde{\Sigma}}(t){\rm e}^{-S(t)}\,.
 \label{ECCM-states}    
\end{equation}
They are the ECCM analogues of the NCCM forms in
eq.~(\ref{NCCM-states}).  The  ECCM form for the
expectation value $\overline{\mathbb{Q}}(t)$ of an arbitrary
physical observable described by the Hermitian operator 
$\mathfrak{q}(t)$ to the NCCM form given by eq.~(\ref{NCCM-ev:a})
is then given in terms of a double similarity
transformation as follows,
\begin{equation}
 \br\mathfrak{q}(t)\kt=\overline{\mathbb{Q}}(t)\equiv
 \br\psi_0|\mathbb{Q}(t)|\psi_0\kt\,;\quad
 \mathbb{Q}(t) \equiv {\rm e}^{\tilde{\Sigma}(t)}Q(t){\rm e}^{-\tilde{\Sigma}(t)}
 ={\rm e}^{\tilde{\Sigma}(t)}{\rm e}^{-S(t)}\mathfrak{q}(t)
 {\rm e}^{S(t)}{\rm e}^{-\tilde{\Sigma}(t)}\,.
 \label{ECCM-ev:a}    
\end{equation}
It is convenient 
to move from the (linked-cluster) set of $c$-number coefficients
$\{s_I(t),\tilde{\sigma}_I(t)\}$ to a new (and also linked-cluster) ECCM set
$\{\sigma_I(t),\tilde{\sigma}_I(t)\}$ belonging to the pair of operators
$\{\Sigma(t),\tilde{\Sigma}(t)\}$, where the new linked operator $\Sigma(t)$
is defined as follows,
\begin{equation}
 \Sigma(t)|\psi_0\kt\equiv(\mathbbm{1}-|\psi_0\kt\br\psi_0|)\,
 {\rm e}^{\tilde{\Sigma}(t)}S(t)|\psi_0\kt=\sum_{I\neq0}\sigma_I(t)C_I^+|\psi_0\kt\,.
 \label{ECCM-Sigma-op}
\end{equation}
The inverse pair of transformations between the coefficients 
$\sigma_I(t)$ and $s_I(t)$
are thus given $\forall I\neq0$ as follows
\begin{equation}
 \sigma_I(t)=\br\psi_0|C_I^-{\rm e}^{\tilde{\Sigma}(t)}S(t)|\psi_0\kt\,,\quad
 s_I(t)=\br\psi_0|C_I^-{\rm e}^{-\tilde{\Sigma}(t)}\Sigma(t)|\psi_0\kt
 \label{ECCM-Sigma-coeffts}
\end{equation}

The ECCM expectation value functional,
$\overline{\mathbb{Q}}(t)=\overline{\mathbb{Q}}
[\sigma_I,\tilde{\sigma}_I;t]=\br\mathfrak{q}(t)\kt$,
of an observable operator
$\mathfrak{q}(t)$ is now considerably more complicated
than its NCCM counterpart $\overline{Q}[s_I,\tilde{s}_I;t]$, due to 
the presence of the double similarity transformation.  A detailed
analysis \cite{Arponen_1983,Arp-Bish-Paj_1987a} has shown, however,
that the ECCM counterparts to the NCCM results of eqs.~(\ref{NCCM-ev:b})
and (\ref{NCCM-ME_L}) are given as follows,
\begin{equation}
 \overline{\mathbb{Q}}[\sigma_I,\tilde{\sigma}_I;t]=\sum_{m=0}^{\infty}\frac{1}{m!}
 \sum_{n=0}^{\infty}\frac{1}{n!}\sum_{J_1\neq0}\cdots\!\sum_{J_m\neq0}
 \sum_{K_1\neq0}\cdots\!\sum_{K_n\neq0}
 \br J_1,\cdots\!,J_m|\mathfrak{q}(t)|K_1,\cdots\!,K_n\kt_{\mbox{\scriptsize{\rm ECCM}}}\,
 \tilde{\sigma}_{J_1}\cdots\tilde{\sigma}_{J_m}\sigma_{K_n}\cdots \sigma_{K_1}\,,
 \label{ECCM-ev:b}    
\end{equation}
where the ECCM matrix elements are now defined as
\begin{equation}
  \br J_1,\cdots\! ,J_m|\mathfrak{q}(t)|K_1,\cdots\! ,K_n\kt_{\mbox{\scriptsize{\rm ECCM}}}
  \equiv \br\psi_0|C_{J_1}^{-}\cdots C_{_m}^{-}\, \mathfrak{q}(t)\,C_{K_1}^+\cdots C_{K_n}^+ 
  |\psi_0\kt_{\mbox{\scriptsize{$\mathcal{DL}$}}}\,.
  \label{ECCM-ME_L}
\end{equation}
The suffix $\mathcal{DL}$ in eq.~(\ref{ECCM-ME_L}) now indicates the ECCM
{\em double linked\,} structure, which
is characterised by the following {\em two\,} constraints: 
(i) from each group of particles
characterised by each of the configuration-space indices $K_j$, 
the amplitude $\sigma_{K_j}$ associated
with the creation operator $C_{K_j}^+$, must have at least one 
particle line connected to the operator $\mathfrak{q}$ 
(exactly as in the $\mathcal{L}$-linking of the NCCM); 
and further (ii) for each group of particles characterised 
by each of the configuration space indices $J_k$,
the amplitude $\tilde{\sigma}_{J_k}$ associated with the destruction 
operator $C_{J_k}^-$  {\em either\,} must have at
least one particle line connected to the operator $\mathfrak{q}$, 
{\em or\,} there must be connections, by at least one particle line 
in each case, to at least {\em two\,} separate amplitudes $\sigma_{K_j}$
associated with two different creation operators $C_{K_j}^+$.

The basic ECCM amplitudes $\{\sigma_I,\tilde{\sigma}_I\}$
may thus be viewed as a set of quasilocal classical fields, due
to the maximal connectivity feature built into the ECCM.
What is meant by quasilocality here is that each of the amplitudes,
which now collectively completely characterize the
theory and exactly describe the ground ket and bra states,
obeys the cluster property in the usual sense of
approaching zero in the limit that any one particle or group
of particles comprising the many-body cluster
becomes far removed from the remainder. 
In turn this permits applications to, e.g., 
topological excitations 
and cases with spontaneous symmetry breaking, that
the NCCM would perhaps have difficulties in describing, due to the
non-locality of the $\{\tilde{s}_I\}$ amplitudes describing the bra states within
the NCCM, which is overcome within the ECCM by the second
similarity transformation.

If we now insert the ECCM wave function parametrizations
of eq.~(\ref{ECCM-states}) into the definition of eq.~(\ref{action-def}) 
of the action functional, and make use of the relationships in
eqs.~(\ref{ECCM-Sigma-op}) and (\ref{ECCM-Sigma-coeffts}), it
is relatively simple to show that the action now takes the ECCM form,
\be
 \mathcal{A}=\int_{t_0}^{t_1} {\rm d}t\,\bigg\{
 -{\rm i}\sum_{I\neq0}\Big(\frac{{\rm d}\tilde{\sigma}_I}{{\rm d}t}\sigma_I\Big)
 -\overline{\mathbb{H}}[\sigma_I,\tilde{\sigma}_I]\bigg\}
 =\int_{t_0}^{t_1} {\rm d}t\,\bigg\{
 {\rm i}\sum_{I\neq0}\Big(\tilde{\sigma}_I\frac{{\rm d}\sigma_I}{{\rm d}t}\Big)
 -\overline{\mathbb{H}}[\sigma_I,\tilde{\sigma}_I]\bigg\}\,.
 \label{ECCM-action}
\ee
The independent (bivariational) stationarity conditions,
$\partial\mathcal{A}/\partial\tilde{\sigma}_I=0=\partial\mathcal{A}/\partial\sigma_I$,
now yield the evolution equations,
\be
 {\rm i}\frac{{\rm d}\sigma_I}{{\rm d}t}
 =\frac{\partial \overline{\mathbb{H}}}{\partial \tilde{\sigma}_I}\,,\quad
 -{\rm i}\frac{{\rm d}\tilde{\sigma}_I}{{\rm d}t}
 =\frac{\partial \overline{\mathbb{H}}}{\partial\sigma_I}\,;
 \quad \forall I\neq 0\,,
 \label{ECCM-evol-eqs}
\ee
for the ECCM cluster correlation coefficients, which are 
the the precise counterparts of eq.~(\ref{NNCCM-evol-eqs})
in the NCCM case.  Equation (\ref{ECCM-evol-eqs}) clearly now shows 
that the ECCM coefficients $\sigma_I$ and $\tilde{\sigma}_I$ are
again a canonically conjugate pair in exactly the same sense
as are $s_I$ and $\tilde{s}_I$ in the NCCM case.  As promised earlier,
we can now see why the operator $S$ was replaced in the ECCM by
the operator $\Sigma$ via eq.~(\ref{ECCM-Sigma-op}).

It is now evident that the NCCM eqs.~(\ref{NCCM_EOM-for-Vbar}) and 
(\ref{NCCM_Poisson-a}) find their completely analogous ECCM 
counterparts,
 \be
 \frac{{\rm d}{\overline{\mathbb{Q}}}}{{\rm d}t}=
 \frac{\partial{\overline{\mathbb{Q}}}}{\partial t}+
 \{\overline{\mathbb{Q}},\overline{\mathbb{H}}\}\,,
 \label{ECCM_EOM-for-Vbar}
 \ee
 \be
 \{\overline{\mathbb{A}},\overline{\mathbb{B}}\}\equiv \frac{1}{\rm i} \sum_{I\neq0}\left(
 \frac{\partial\overline{\mathbb{A}}}{\partial\sigma_I}\,
 \frac{\partial\overline{\mathbb{B}}}{\partial\tilde{\sigma}_I}-
 \frac{\partial\overline{\mathbb{A}}}{\partial\tilde{\sigma}_I}\,
 \frac{\partial\overline{\mathbb{B}}}{\partial\sigma_I}
 \right)\,.
 \label{ECCM_Poisson-a}
 \ee
The corresponding ECCM expresion for the expectation value
$\overline{\mathbb{AB}}$ of the product of two ECCM (doubly 
similarity-transformed) operators is now considerably more
complicated than its NCCM counterpart in eq.~(\ref{exp-value_for-product-AB}).
Nevertheless, it can be shown \cite{Arp-Bish-Paj_1987a} that the
expectation value of the commutator of operators $\mathbb{A}$ and
$\mathbb{B}$ takes the form
 \be
 \br[\mathfrak{a},\mathfrak{b}]\kt=
 \overline{[\mathbb{A},\mathbb{B}]}=
 \overline{\mathbb{AB}}-\overline{\mathbb{BA}}=
 {\rm i}\{\overline{\mathbb{A}},\overline{\mathbb{B}}\}\,,
 \label{ECCM_commutator-map-to-Poisson}
 \ee
within the ECCM, which is the exact analogue of its NCCM
counterpart in eq.~(\ref{NCCM_commutator-map-to-Poisson}).
It should by now be evident that there are many other analogies
between the NCCM and ECCM, which we do not now enumerate further
however.

In conclusion, we reiterate that both versions of the CCM can,
as we have shown, be formulated in terms of 
a variational (or, more properly, a bivariational) principle 
for either the stationary (i.e., time-independent) or time-dependent
Schr\"{o}dinger equations \cite{Arponen_1983,Bish-Arp-Paj_1989}.  
For the stationary
(S-CCM) cases of both the NCCM and ECCM  
the bivariational principle is 
for the ground-state expectation value functional
of the Hamiltonian, while for the time-dependent 
(TD-CCM) cases it is for the
action functional.

\subsection{Illustration of stationary aspects of the CCM: the Rabi model \label{Raa-sub}}

We will elaborate further on the CCM 
in sect.~\ref{uctvrta} where we discuss the 
physics that lies behind the introduction of
non-Hermitian operators, after presenting more of the mathematical
framework. For now, however, let us elucidate the use of the
CCM in practice via a simple, yet illustrative, example. 
Among a plethora of possible applications, we 
choose the Rabi model, an important and archetypal 
model in quantum optics where it describes a two-level atom (with
energy spacing $\hbar\omega_0$)
coupled via a dipole interaction (of coupling strength proportional
to $g$) to a single mode 
of a quantized electromagnetic radiation
field (with frequency $\omega$) \cite{Allen-Eberly-book_1978}. In units
where $\hbar=1$, the model is defined by 
the Hamiltonian,
\begin{equation}
 \mathfrak{h}= \frac{1}{2} \omega_{0} \sigma^z + \omega b^{\dagger}b + g \left(\sigma^{+} + \sigma^{-}\right)
       \left(b^{\dagger} + b \right) \,,          
 \label{Rabi-Hamiltonian}
\end{equation}
in terms of pseudo-spins
\begin{equation}
\sigma^z=
\begin{pmatrix}
1 & 0\\
0 & -1
\end{pmatrix}\,,\quad
\sigma^+=
\begin{pmatrix}
0 & 2\\
0 & 0
\end{pmatrix}\,,\quad
\sigma^-=
\begin{pmatrix}
0 & 0\\
2 & 0
\end{pmatrix}\,,
\end{equation}
related to the conventional Pauli matrices,
\begin{equation}
\sigma^x = \frac{1}{2}\left(\sigma^+ + \sigma^-\right),~~ \sigma^y =
\frac{\rm i}{2}\left(\sigma^- - \sigma^+\right).
\end{equation}
The field mode is described in terms of the annihilation and creation operators, 
$b$ and $b^\dagger$,
respectively, which obey the usual bosonic commutation relation,
\begin{equation}
\left[b,b^{\dagger}\right] = \mathbbm{1}.
\end{equation}
Within the context of nuclear magnetic resonance, the 
model also mimics a spin interacting with a 
field of phonons \cite{Rabi_1937}. A close relationship of the model
also exists to the static Lee model of nuclear interaction
\cite{Marshall-Pell_1981}.

From our present viewpoint the Hamiltonian of eq.~(\ref{Rabi-Hamiltonian}) is
suitable not only as a rather transparent model of physical
reality but also as a useful example to illustrate the
applicability and efficiency of the standard NCCM
approach (and see refs.~\cite{Bish-Dav-Qu-vdW_1996,Bishop-Emary_2001} for more details).
A convenient choice of reference state for this model is now $|\psi_0\kt=
|0\kt|\!\downarrow\kt$, which is just the ground state of the system for $g=0$, 
assuming $\omega_0>0$ (viz., an empty field mode and an unexcited atom), 
in a notation where the first ket refers to zero bosons in the 
occupation number representation and the second to the lower state of the two-level
atom in an obvious pseudo-spin representation with the spin quantized in the 
$z$-direction.  For example, the generic NCCM correlation operators of 
eq.~(\ref{NCCM-ops}) can now be written in the following specific forms,
\begin{subequations}
 \be
 S=S_1 +S_2\,;\quad S_1=\sum_{n=1}^{\infty}s_n^{(1)}(n!)^{-1/2}(b^\dagger)^n\,, \quad
 S_2=\sum_{n=1}^{\infty}s_n^{(2)}[4(n-1)!]^{-1/2}(b^\dagger)^{n-1}\sigma^+\,,
 \label{NCCM-op-S:Rabi}
 \ee
 \be
 \tilde{S}=\mathbbm{1}+\tilde{S}_1 +\tilde{S}_2\,;\quad
 \tilde{S}_1=\sum_{n=1}^{\infty}\tilde{s}_n^{(1)}(n!)^{-1/2}b^n\,, \quad
 \tilde{S}_2=\sum_{n=1}^{\infty}\tilde{s}_n^{(2)}[4(n-1)!]^{-1/2}b^{n-1}\sigma^-\,.
 \label{NCCM-op-Stilde:Rabi}
 \ee
\end{subequations}
The only approximation that is now made is to truncate the sums in 
eqs.~(\ref{NCCM-op-S:Rabi}) and (\ref{NCCM-op-Stilde:Rabi}) at the term
with $n = N$, giving the so-called SUB-$N$ approximation, 
in which all the coefficients $s_n^{(i)}$ and
$\tilde{s}_n^{(i)}$, $i=1,2$, are set to zero $\forall n > N$. 

In the stationary version of the NCCM, one may now readily evaluate
the ground-state expectation value, 
$\br\mathfrak{q}\kt=\overline{Q} = \overline{Q}[s_I,\tilde{s}_I]$
of any physical observable $\mathfrak{q}$ for the
Rabi model, as described in sec.~\ref{NCCM-fund}. For example, 
the ground-state energy, $E=\br\mathfrak{h}\kt$, has been so calculated
at various NCCM SUB-$N$ levels of approximation in ref.~\cite{Bish-Dav-Qu-vdW_1996},
to which the interested reader is referred for further details. We note
only that with the above choices of reference state and cluster operators the
NCCM was found to yield a (probably spurious) phase transition at
a value $g \to g_c \approx 0.665$ of the coupling constant, which signalled
a breakdown of the (convergence of the) calculation for values $g>g_c$, 
which is precisely the region where the ground state becomes nearly
degenerate in energy with the first excited state of the system \cite{Emary-Bishop_2002}.
Still, the method provided excellent results for $E$ in the region 
$g<g_c$.  We defer further discussion of applications of the TD-CCM to this
and other systems to sect.~\ref{models}.

\section{Excited states and  non-Hermitian reformulations of conventional quantum mechanics\label{druha}}

We shall consider the question of how to construct excited states
and the excitation spectrum of a quantum many-body system within 
the CCM later in sec.~\ref{ctvrta}.  However, we note now that
whenever one tries to move beyond the CCM-based precise  evaluations of the  
ground-state characteristics of a generic (i.e., not even necessarily just many-body) 
quantum system, one may encounter a number of new, 
and sometimes unexpected,  methodical  challenges. 
At an initial level of approach it appears that one must find a way of making 
any CCM-inspired explicit construction of an optimal Dyson map  $\Omega =\exp S$ of
eq.~(\ref{expS})
sufficiently 
less reference-dependent.
For example,
in the nuclear-physics IBM-related setting of ref.~\cite{Holstein-Primakoff_1940}, 
this appeared to be a decisive technical obstacle. 

The essence of the 
difficulty lies in the fact that the descriptive ambitions of the
IBM 
constructions 
involved not only the  
single ground state but also whole multiplets of the low-lying excited states.
In this setting, a decisive amendment of the variational results 
(see the thorough review and mathematical clarification and explanation of 
Scholtz et al~\cite{Scholtz-Geyer-Hahne_1992})
only appeared after a return of attention 
to the older Dyson's papers on ferromagnetism
 \cite{Dyson_1956a,Dyson_1956b}
in which calculational success 
was essentially based on the use of a  
{\em non-unitary\,} operator $\Omega$ in eq.~(\ref{fakto}).

The latter idea has later been identified as yielding a nontrivial form for the auxiliary 
operator  product $\Omega^\dagger\Omega$ which is henceforth called the {\em metric operator\,} 
of the physical Hilbert space,
 \be
 \Omega^\dagger\Omega \equiv \Theta \neq \mathbbm{1}\,,
 \label{product}
 \ee
where $\mathbbm{1}$ is again the identity operator.
During the subsequent developments
of the field 
different authors   
succeeded 
in finding techniques 
circumventing the obstacles 
emerging due to the nontrivial nature of the metric (i.e., that 
it is not simply the identity operator), which is not
encountered in conventional Hermitian quantum mechanics.


The computational economy of the Hermitian  
Schr\"{o}dinger-picture (SP) formulation  
of quantum mechanics \cite{Schroedinger_1926} is 
very persuasive. It is just this aspect of the 
formulation that 
explains its popularity and success 
in many branches of quantum physics and quantum chemistry.
Nevertheless, several existing technical limitations
of the conventional SP description 
of quantum mechanics have almost 
always forced its users to search for amendments.
In this context a short letter by Bender and Boettcher 
\cite{Bender-Boettcher_1998} 
proved to be very influential in 
returning attention to 
the less usual 
but fully admissible 
possibility of non-Hermitian representations of observables as promoted, a few years earlier, 
in ref.~\cite{Scholtz-Geyer-Hahne_1992}.
The latter paper itself actually  recalled the older idea of
mapping certain Hermitian fermion operators onto simpler non-Hermitian 
boson operators by means of the generalized Dyson 
mapping \cite{Janssen-et-al_1971}, with all of these developments being  
precisely the motivation behind the IBM technique
itself
\cite{Arima-Iachello_1976,Arima-Iachello_1978,Arima-Iachello_1979}.

Such a ``crypto-Hermitian'' 
amendment (or perhaps, better, extension)
of our understanding of quantum theory has, at present, multiple 
parallels and continuations \cite{Mostafazadeh-ijgmmp_2010}.
The resulting, truly deep, theoretical 
reconsideration of the first principles of quantum theory
has now become widely accepted by a broad 
community of physicists. The impact of the idea may nowadays be detected even 
beyond the domain of quantum 
theory \cite{Znojil_in-Bagarello-et-al-book_2015}.
Still, its origins may be dated back to the studies of
certain truly complicated many-fermion quantum systems.
It was, in fact, Dyson \cite{Dyson_1956a,Dyson_1956b}
(and, independently, Maleev \cite{Maleev_1957})
who proposed, more than sixty years ago, and mainly for the 
purely practical purposes of variational calculations,
a key idea lying in the (formally reversible) {\em non-unitary} mapping
of wave functions as in eq.~(\ref{fakto}).

The reader should be aware at this point that up till now we have not
precisely defined the term ``Hermitian'' on the mathematical level.
In general, one would need to distinguish with care
such alternative concepts related to the term ``Hermitian'' as ``symmetric'',
``self-adjoint'' (on a specified domain), and ``essentially self-adjoint'' (with
a well-defined core). However, in most of the prospective applications that we have 
in mind we work with a finite-dimensional matrix form of the so-called
``Hermitian'' operator, in which case there is no longer any need for such
mathematical precision. 
Marginally, let us note here that, even in the infinite-dimensional and non-matrix models,
the essence of many related apparent paradoxes can be identified as lying in the fact that
the physical Hilbert space is exclusively presented 
via its representation in ${\cal H}^{\rm (user\!-\!friendly)}_{[\rm unphysical]}$.
Thus, the
clarification of the Hermiticity/non-Hermiticity misunderstandings 
becomes simple when one consequently stays in the latter space and
characterizes the switch to  ${\cal H}^{(\rm final)}_{[\rm physical]}$
by the mere change of the inner product (and the interested reader can find further 
useful comments on this trick in  ref.~\cite{Mostafazadeh-ijgmmp_2010}).

Due to the latter trick, 
phenomenolgical Hamiltonians   may be, admittedly,
manifestly non-Hermitian in a preselected and, presumably, just
unphysical, auxiliary Hilbert space  
${\cal H}^{(\rm user\!-\!friendly)}_{[\rm unphysical]}$.
As a concrete example
let us recall the most popular, 
ordinary differential, upper-case Hamiltonian
 $
 H^{(\rm imaginary \ cubic)}=-\frac{{\rm d}^2}{{\rm d}x^2} + {\rm i} x^3\,,
 \label{icu}
 $
which is
${\normalsize \cal PT}$-symmetric \cite{Bender-rpp_2007,Bender-Boettcher_1998}
but manifestly non-Hermitian in ${\cal H}^{(\rm user\!-\!friendly)}_{[\rm unphysical]}
= L^2(\mathbb{R})$.
After an appropriate amendment of the inner product, according to Bender \cite{Bender-rpp_2007},
the underlying quantum system ${\cal S}$ may be assigned
its conventional quantum probabilistic unitary-evolution interpretation
via the reconstruction of the physical
Hermitian conjugation [c.f., the ``physical'' conjugation 
$H \to H^\ddagger \equiv \Theta^{-1}{{H}}^\dagger \Theta$ 
in place of the conventional, friendlier but unphysical, 
conjugation $H \to H^\dagger$.


There exist many interesting aspects of the
model described by $H^{(\rm imaginary \ cubic)}$
and by its various
alternatives, which are explained and discussed
in, e.g., the
reviews presented in refs.~\cite{Bagarello-et-al-book_2015,Bender-rpp_2007,Mostafazadeh-ijgmmp_2010}.
For all of them the underlying
probabilistic interpretation of the
quantum systems is based on the innovative use of
non-Hermitian Hamiltonian avatars $H$ with real energy eigenvalue spectra.
Almost without exception, all of these examples
are presented,  in the current literature,
in the framework of what we will call here the generalized (i.e.,
``non-Hermitian'')
Schr\"{o}dinger-picture formalism. In its stationary case we will call it, 
for the sake of definiteness, the Dyson-Maleev formalism (DMF), 
for a particularly extensive review of the generalized forms of
which we may particularly recommend ref.~\cite{Janssen-et-al_1971}.

Once one manages to achieve
at least a reasonable degree of reference-independence
of $\Omega$, a suitable combination  of the 
CCM mathematics with IBM physics could open new 
construction horizons.
Indeed, in the strictly stationary case, 
i.e., in the case with  property 
 $
 \Omega(t) = 
 \Omega(0)\,,
  \forall t$,
there exists an intimate connection and correspondence 
between the generality (i.e., non-unitarity) of $\Omega$,
the non-Hermiticity of the avatars $H$ of the Hamiltonians
given by eq.~(\ref{obshamst})
and the bivariational nature of the CCM recipes, as we discussed 
in our previous paper \cite{Bishop-Znojil_2014}.

The DMF 
approach
may be perceived as one of the standard numerical algorithms,
which
transfers the description 
of the states from the traditional 
(i.e., often, fermionic, 
Fock)
Hilbert space 
 $
 {\cal H}^{(\rm initial)}_{(\rm DMF)}={\cal H}^{(\rm fermionic)}_{(\rm DMF)}
 $
of wave functions $\psi$
to its {\em formally non-equivalent\,}
 (i.e., often, effective, bosonic)  Hilbert space
${\cal H}^{(\rm unphysical)}_{(\rm DMF)}={\cal H}^{(\rm bosonic)}_{(\rm DMF)}$
of wave functions $\psi_0$.
The resulting gain in flexibility is 
remarkable, being broadly appreciated and widely applied \cite{Arima-Iachello_1976,Arima-Iachello_1978,Arima-Iachello_1979,Janssen-et-al_1971}.
In the context of the DMF theory 
the non-equivalence
of the two Hilbert spaces ${\cal H}^{(\rm fermionic)}_{(\rm DMF)}$
and ${\cal H}^{(\rm bosonic)}_{(\rm DMF)}$ proved inessential.
In the latter space 
one can always amend the inner product
in such a way that the resulting new,
third  Hilbert space 
${\cal H}^{(\rm final)}_{(\rm DMF)}$ becomes eligible to
play the role of the ultimate Dyson-Maleev physical Hilbert space~\cite{Dyson_1956a,Maleev_1957,Janssen-et-al_1971,Holstein-Primakoff_1940,Beliaev-Zelevinsky_1962,Schwinger_1952,Marumori-et-al_1964,Sorensen_1967}.
For this reason the predictions of the DMF approach are essentially
identical  to those of the conventional  Schr\"{o}dinger picture, provided only that
the third, final space is, by construction, assumed unitarily equivalent to the initial 
one, 
$
{\cal H}^{(\rm final)} \sim {\cal H}^{(\rm initial)}
$.
For this reason
the key features
of the stationary DMF scheme (typically, in its nuclear-physics IBM
implementations)
remain transparent. Its structure
may be summarized by the following compact illustrative
flowchart diagram,
{\normalsize
 \be
 \ \ \ \ \ \ \ \ \
 \ \ \ \ \ \ \ \ \ 
 \ \ \ \ \ \ \ \ \ 
 \ba
  \\
    \ \ \ \ \ \ \ \ \begin{array}{|c|}
 \hline
 \\
   \fbox{\rm DMF\ input\ information}  \\ 
   \\
   {\rm realistic\ } {\rm self\!\!-\!\!adjoint\ Hamiltonian\ }
   {{\mathfrak{h}}}\
  {\rm }\\
      {\rm {\bf physical} {\   (fermionic,\ } Fock)\  space\ }
    {\cal H}^{(\rm initial)}_{[\rm textbook]}\\
    \\
 \hline
 \ea
  \ \ \ \ \ \ \ \ \ \ \  \  \  \ \ \ \ \ \
  \ \ \ \ \ \ \ \ \ \ \ \ \ \ \ \ \ \ \  \  \  \ \ \ \ \
  \ \ \ \ \ \ \ \
 \\
 \\
 \stackrel{{\bf Hilbert-space\  map} \  \Omega^{-1}\, }{}
 \ \
  \swarrow\ \  \ \ \ \ \ \ \ \ \ \ \ \ \ \ \ \ \ \ \ \ \ \ \ \
  \ \ \ \ \ \
 \ \ \ \ \  \searrow \nwarrow\
 \stackrel{\bf   equivalence}{} \ \ \ \ \ \ \ \
  \ \ \  \  \  \ \ \ \ \ \ \ \ \ \ \ \ \ \ \ \
  \ \ \ \ \ \ \ \ \ \ \ \ \ \ \ \ \ \ \  \  
  \ \ \ \ \ \ \ \ \\
  \\
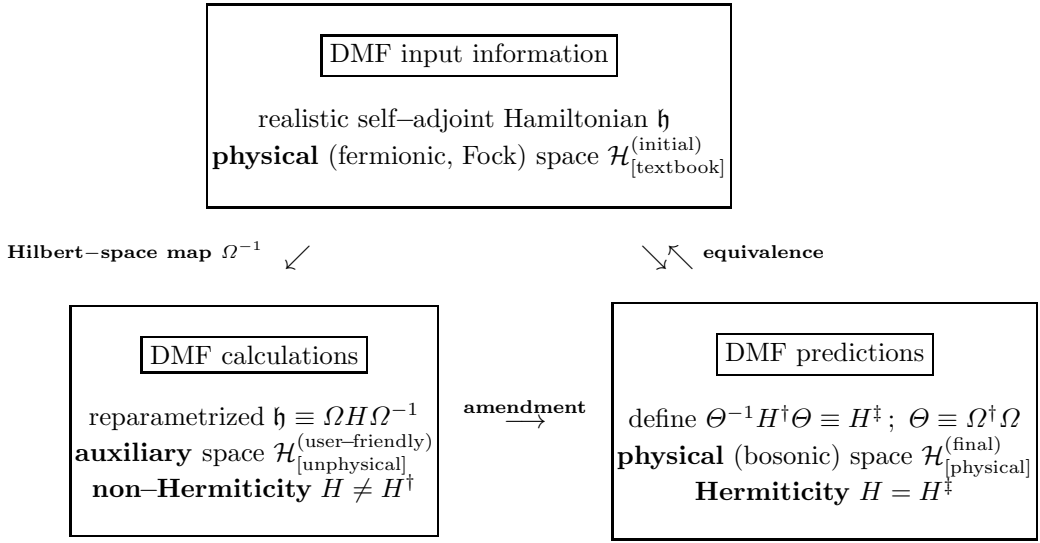
 \begin{array}{|c|}
 \hline
 \\
   \fbox{\rm DMF\ calculations} \\
   \\
    {\rm  reparametrized}\  {{\mathfrak{h}}}
   \equiv \Omega{{H}}\Omega^{-1}\  \\
    {\rm   {\bf  auxiliary}  \ space\  }
    {\cal H}^{(\rm user\!-\!friendly)}_{[\rm unphysical]} \\
   {\rm  \bf non\!\!-\!\!Hermiticity\ }{{{H}}}
    \neq {{{H}}}^\dagger\\
    \\
  \hline
 \end{array}
 \stackrel{  \ \ {\bf amendment}  \ \  }{ \longrightarrow }
 \begin{array}{|c|}
 \hline
 \\
   \fbox{\rm DMF\ predictions} \\
   \\
 {\rm define}\ \Theta^{-1}{{H}}^\dagger\Theta \equiv {{{H}}}^\ddagger
 \,; \,\, \Theta \equiv \Omega^{\dag} \Omega   \\
 {\rm  { \bf physical}\   ( bosonic) \ space}\ {\cal H}^{(\rm final)}_{[\rm physical]}\
     \\
   {\rm  \bf Hermiticity\ }{{{H}}}
   = {{{H}}}^\ddagger\\
     \\
 \hline
 \ea \ \ \ \ \ \ \ \ \ 
 \ \ \ \ \ \ \ \ \ \ 
 \ \ \ \ \ \ \ 
  \\
  \ \ \ \\
    \end{array}
 \label{THSstac}
 \ee
One of the most remarkable consequences of
the non-unitarity of the mapping $\Omega$, seen
in such a diagram lies in the 
{\em coexistence\,} of the
Hermiticity and non-Hermiticity properties of
{\em the same\,} upper-case operator $H$, depending on which
Hilbert space one is considering.
In the stationary setting the above-mentioned operator $H$ 
represents an observable quantity, {\em despite} being
non-Hermitian (in the sense that $H \neq H^\dagger$)
in  ${\cal H}^{(\rm user-friendly)}_{[\rm unphysical]}$,
precisely {\em because} it is
self-adjoint (in the sense that $H=H^\ddagger$)
in ${\cal H}^{(\rm final)}_{[\rm physical]}$.

\subsection{Transition to non-stationary quantum systems\label{tdt}}

Our present study 
was inspired by the recent progress in the 
development of what could be called reference-independent
IBM-like theories 
\cite{Znojil_SIGMA_2009,Znojil_PRD_2008,Znojil_ijtp_2013,Znojil-pla_2015,Znojil_AnnPhys_2017}.
The message of our present paper
should be seen in the transfer of 
our current understanding of the merits of the mappings 
of eq.~(\ref{fakto}) 
from the IBM-related context into the bivariational
constructive CCM strategies. 
The NHIP formulation of quantum mechanics will be 
used, in such an application, in a slightly narrower sense, being reserved to cover only
the picture of reality
with the {non-stationarity} property, 
 $$
 \Omega_{(\rm NHIP)} = \Omega_{(\rm NHIP)}(t)\,.
 $$
According to the first consistent introduction of
the full-fledged non-stationary NHIP formalism in
ref.~\cite{Znojil_PRD_2008},
its basic idea may be perceived as a time-dependent extension
of the old IBM-like variational recipe. While its innovated form
is definitely  more flexible, it is undoubtedly
also much more complicated technically \cite{Znojil_ijtp_2013}.
For this reason it is, therefore, perhaps
not too surprising that
the TD-CCM/NHIP relationship has not yet been studied. It
is precisely this omission that we aim to remedy here.
This technical complexity also probably explains why only
a relatively few realistic applications 
of several alternative implementations of the NHIP ideas
themselves may yet be found in the current literature 
\cite{Znojil_AnnPhys_2017,Znojil_ijtp_2013,Faria_Fring_2006,Faria_Fring_2007,Bila-phdthesis_2008,Bila-arXiv_2009,Gong-Wang_2010,Gong-Wang_2013,Maamache_2015,Luiz-Pontes-Moussa-arXiv_2016,Khantoul-Bounames-Maamache_2017}.

%
%

Since the birth of quantum theory in its Heisenberg-picture 
(HP) formulation \cite{Heisenberg_1925} 
and, less than a year later, in its
Schr\"{o}dinger-picture (SP) formulation \cite{Schroedinger_1926},
those attempting to apply the theory have
always needed to resolve the conflict
between the more intuitive nature of the HP quantization
of observables and
the maximal economy of the transfer of attention to the wave functions
in the SP approach. A partial relief of this
tension came, later, with the invention of the more universal
interaction-picture (IP) (alias the Dirac-picture) formalism  from which
the SP and HP descriptions of quantum dynamics could have been deduced
as two separate special limiting cases.
An enhanced flexibility of the ensuing 
language then also gave rise to the well-known 
successes of the manifold 
IP applications (and predictions) 
in practice, especially in the context of
quantum field theory and quantum many-body theory 
\cite{Bjorken-Drell-book_1965,Negele-Orland-book_1998}.

In table~\ref{dowe}
we present a comparison of the HP, SP and IP
``strictly Hermitian'' descriptions of unitary (i.e., stable)
quantum evolution. As a comment on the table we might
emphasize that the same physics is described by the
{\em  single\,} operator
evolution equation in the HP formulation, and by the {\em  single\,}
ket-vector evolution
equation in the SP formulation, as well as 
by a {\em pair\,} of evolution equations in
the IP formulation. For a compensation
of the seeming disadvantage of the doubling
of the number of evolution equations in the IP formalism, 
we note again that the latter IP picture contains both the
former (HP and SP) ones as special limiting cases.
\begin{table}[hb]
\caption{Conventional Hermitian-generator descriptions of quantum dynamics }
 \label{dowe}
\centering
\begin{tabular}{||c||c|c||}
\hline \hline 
  \multicolumn{1}{||c||}{\rm picture}
   &{\rm  observables} $\mathfrak{q}(t)$ &{\rm state vectors} $\psi(t)$
     \\
 \hline
 \hline
  {\rm HP} &{\rm evolving}
  &{\rm constant}\\
 \hline
  {\rm SP} &{\rm constant}
  &{\rm evolving} \\
 \hline 
  {\rm IP} &{\rm evolving}
  &{\rm evolving} \\
 \hline \hline
\end{tabular}
\end{table}


The ultimate choice between the HP, SP and IP
(or, indeed, many other \cite{Styer_et-al_2002}) 
model-building strategies depends, first of all,
on the actual form
of our specification of the quantum system in question.
For this reason,
most standard textbooks usually prefer the SP language, 
only adding the HP and IP
analyses of quantum dynamics 
at the later stages of explanation.
This makes the SP-based
specification of
quantum dynamics less intuitive but shorter, based on the 
rather formal introduction of
a ``physical''
Hilbert space ${\cal H}^{\rm{(initial)}}$
and of a suitable
self-adjoint ``Hamiltonian'' $\mathfrak{h}=\mathfrak{h}^\dagger$
defined within that space.
Many researchers prefer 
the use of the SP language in practice, since it
requires, in addition, a
maximally realistic origin for,
and ``derivation'' of, the
latter Hamiltonian operator, which is typically found, e.g.,
via a ``quantization'' of its
suitable classical-physics counterpart.


\subsection{Non-Hermitian versions of the Heisenberg and Dirac pictures \label{NHHP}}

The key to the extension of the validity
of the IBM-type pattern beyond its stationary DMF
version of eq.~(\ref{THSstac})
has been found
in ref.~\cite{Znojil_PRD_2008}.
First of all, the removal of the existing theoretical obstacles 
and objections (and see, e.g., ref.~\cite{Mostafazadeh-plb_2007}) required
a refinement of the terminology.
In ref.~\cite{Znojil_PRD_2008} it has been emphasized that the
initial and observable physical
Hamiltonian $\mathfrak{h}(t)$
(defined in ${\cal H}^{(\rm initial)}_{[\rm  textbook]}$; 
in general it may be time-dependent)
merely becomes replaced, as in eq.~(\ref{obslambda}), by its upper-case
isospectral {\em non-stationary} avatar, 
 \be
 H(t) \equiv \Omega^{-1}(t)\mathfrak{h}(t)\Omega(t)\,,
 \label{obsham}
 \ee
which is non-Hermitian in  ${\cal H}^{(\rm user-friendly)}_{[\rm unphysical]}$
but Hermitian in ${\cal H}^{(\rm final)}_{[\rm physical]}$ 
[i.e., in the sense that $H=H^{\ddagger}$, where
$H^{\ddagger} \equiv \Theta^{-1}{{H}}^\dagger\Theta$, as in
the flowchart of eq.~(\ref{THSstac})]. 
For this reason,
both of the operators $\mathfrak{h}(t)$ and $H(t)$
represent an instantaneous energy, i.e.,
the same observable physical quantity. 
Clearly, eq.~(\ref{obsham}) is simply the time-dependent
counterpart of its stationary equivalent in 
eq.~(\ref{obshamst}).
However, the clarification of the 
{\em dynamical role} of the two operators
$\mathfrak{h}(t)$ and $H(t)$ in eq.~(\ref{obsham}) has
turned out to be much less straightforward.
In the literature, 
the process of this clarification was both lengthy and tedious
\cite{Mostafazadeh-plb_2007,Znojil_PRD_2008,Znojil-arXiv_2007a,Mostafazadeh-arXiv_2007a,Znojil-arXiv_2007b,Znojil-arXiv_2007c,Mostafazadeh-arXiv_2007b}.
Fortunately, 
at the end of this process 
in 2009
(c.f., ref.\ \cite{Znojil_SIGMA_2009} and some later 
addenda in refs.\ \cite{Gong-Wang_2010,Gong-Wang_2013,Maamache_2015})
the eventual outcome has transpired to be both 
relatively elementary
and transparent, as we now briefly explain. 

Its brief summary may start from a return to
table~\ref{dowe}, which reminds us that even in
the conventional Hermitian SP 
setting the SP\,$\to$\,HP transition may be perceived as
mediated by a mapping of the form of 
eq.~(\ref{fakto}), but in which the operator $\Omega$ would be
unitary but manifestly time-dependent.
In 2007, the
feasibility of the extension
of the SP\,$\leftrightarrow$\,HP correspondence to
non-Hermitian cases
was opposed by
Mostafazadeh \cite{Mostafazadeh-ijgmmp_2010,Mostafazadeh-plb_2007,Mostafazadeh-arXiv_2007a,Mostafazadeh-arXiv_2007b}.
Fortunately,
the apparently insurmountable obstacles
and obstructions that were initially raised against the
free applicability of the non-stationarity 
postulate, $$\Omega_{(\rm NHHP)} = \Omega_{(\rm NHHP)}(t)\,,$$ thereafter
appeared in essence to be of a purely terminological 
nature \cite{Znojil_PRD_2008,Znojil-arXiv_2007a,Znojil-arXiv_2007b,Znojil-arXiv_2007c,Fring-Moussa_2016a,Fring-Moussa_2016b}.
As a consequence, the transition to the generalized
``non-Hermitian'' HP (viz., the NHHP)
was eventually formulated definitively in ref.~\cite{Znojil-pla_2015}.
Soon thereafter it was also found to be both feasible and useful 
in practice in some specific applications \cite{Miao-Xu_2016}.


Once we are given a lower-case SP operator $\mathfrak{q}$
representing
an arbitrary observable quantity,
the only relevant task for theorists is a
prediction  of the results of experiments based on the evaluation
of its expectation value.
This means that we have to evaluate the quantity 
$$\bbr \psi(t)|Q(t)|\psi(t)\kt\,.$$
The precise meaning of the ket state $|\psi(t)\kt$
and of the bra (or, rather, brabra) 
state $\bbr \psi(t)|$, are to be specified of course 
(see sect.~\ref{recommended-notation} below).
The upper-case symbol $Q(t)$
just represents  here the operator  (isospectral to $\mathfrak{q}$), 
which is defined in  
the space ${\cal H}^{(\rm user\!-\!friendly)}_{[\rm unphysical]}$ 
via a similarity transformation
 \be
 Q(t) \equiv \Omega^{-1}(t) \mathfrak{q} \Omega(t)\,,
 \label{obslambda}
 \ee
in complete analogy to its NCCM counterpart in
eq.~(\ref{CCM-gen-sim-Xm}).
The overall theoretical flowchart then
has the following compact form,

{\normalsize
 \be
   .  \ \ \ \ \ \ \ \ 
   \ba
  \\
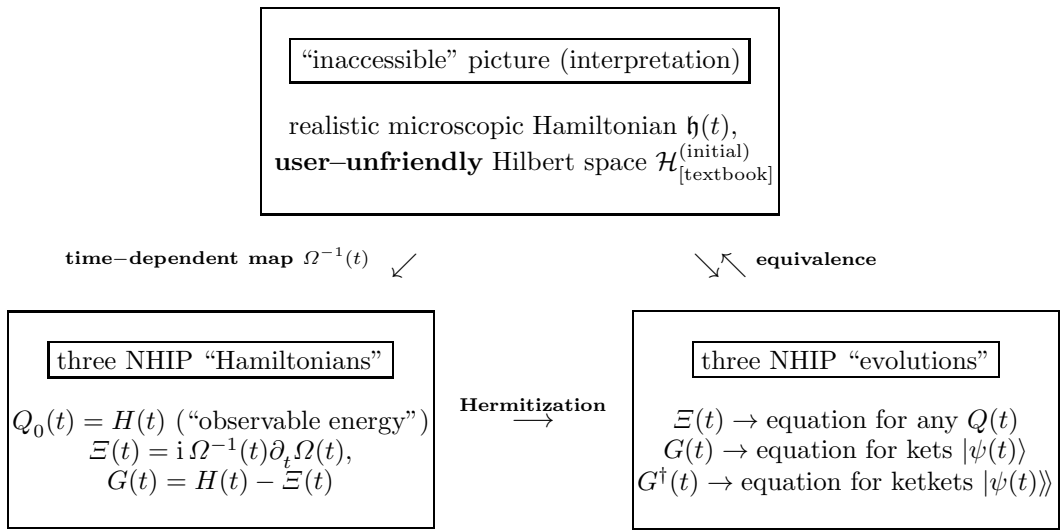
  \begin{array}{|c|}
 \hline
 \\
   \fbox{\rm ``inaccessible"\  picture\ (interpretation)} \\
   \\
   {\rm realistic\ } {\rm  microscopic\ Hamiltonian\ }
    {{\mathfrak{h}}}(t),\
  {\rm }\\
      {\rm  {\ \bf user\!\!-\!\!unfriendly\ }Hilbert\  space\ }
    {\cal H}^{(\rm initial)}_{[\rm textbook]}\\
    \\
 \hline
 \ea
  \ \ \ \ \ \ \ \ \ \ \  \  \  \ \ \ \ \ \ \ \ \
  \ \ \ \ \ \ \ \ \ \ \ \ \ \ \ \ \ \ \  \  \  
  \ \ \ \ \ \ \ \
 \\
 \\
  %
 \stackrel{{\bf time-dependent\  map} \  \Omega^{-1}(t)\, }{}
 \ \
  \swarrow\ \  \ \ \ \ \ \ \ \ \ \ \ \ \ \ \ \ \ \ \ \ \ \ \ \
 \ \ \ \ \  \searrow \nwarrow\
 \stackrel{\bf   equivalence}{} \ \ \ \ \ \ \ \
  \ \ \  \  \  \ \ \ \ \ \ \ \ \ \ \ \ \ \
  \ \ \ \ \ \ \ \ \ \ \ \ \ \ \ \ \ \ \  \  \  \ \ \ \ \ \
  \ \ \ \ \ \ \ \ \\
  \\
 \begin{array}{|c|}
 \hline
 \\
   \fbox{\rm three\  NHIP\ ``Hamiltonians''}\\
   \\
Q_0(t)=H(t)\ {\rm (``observable\ energy")}\\
  {{{\Xi(t)}}}
    =  {\rm i}\,\Omega^{-1}(t)\partial_t\Omega(t), \\
   {{{G(t)}}}
    = H(t)-\Xi(t)  \\
   \\
  \hline
 \end{array}
 \stackrel{  \ \ {\bf  Hermitization}  \ \  }{ \longrightarrow }
 \begin{array}{|c|}
 \hline
 \\
   \fbox{\rm three\  NHIP\ ``evolutions"}\\
   \\
   {{{\Xi(t)}}} \to {\rm equation\ for\ any\ }Q(t)
  \\
   G(t) \to  {\rm equation\ for\  kets} \  |\psi(t)\kt\
     \\
   G^\dagger(t) \to  {\rm equation\  for\  ketkets} \    |\psi(t)\kkt
     \\
     \\
 \hline
 \ea \ \ \ \ \ \ \ \ \ \ \  \  \  \ \ \ \ \ \
  \ \ \ \ \ \ \ \ \ \ \ \ \ \ \ \ \ \ \  \  \  \ \ \ \ \ \
  \ \ \\
  \\
   \end{array}
 \vspace{0.4cm}  
 \label{THSecht}
 \ee
}
\hspace{-0.2cm} 
The precise meaning and definitions of the operators
and state vectors in eq.~(\ref{THSecht}) is given below in 
sect.~\ref{NHevolu-eqns}.
On this basis we may expect that the 
use of the alternative formulations 
of the dynamical equations will 
not hide
their one-to-one correspondences with those in the Hilbert space
$ {\cal H}^{(\rm initial)}_{[\rm textbook]}$ of the
topmost-box in the diagram comprising eq.~(\ref{THSecht}).


\section{Non-Hermitian versions of dynamical 
evolution equations\label{NHevolu-eqns} }

Curiously enough,
the above-outlined NHHP formalism is ``quasi-stationary''
because the HP metric itself remains time-independent  \cite{Znojil-pla_2015,Znojil_ijtp_2013},
 \be
 \Theta_{(\rm HP)}(t) = \Theta_{(\rm HP)}(0)\,.
 \label{nestac}
 \ee
As a consequence, it was necessary to move beyond the constraint (\ref{nestac})
(c.f., ref.~\cite{Znojil_AnnPhys_2017}). 
The current state of the art
is summarized in table \ref{trewe}
in which the third ``non-Hermitian'',
fully general formalism of refs.~\cite{Znojil_AnnPhys_2017,Znojil_PRD_2008}
is now assigned the abbreviation NHIP. Clearly, table~\ref{trewe} is just the general counterpart
in our non-Hermitian context of the earlier table~\ref{dowe} that
pertains to the conventional Hermitian-generator description of
unitary evolution.


\begin{table}[hbt]
\caption{Unitary evolution in non-Hermitian pictures}
 \label{trewe}
 \centering
\begin{tabular}{||c||c|c|c||}
\hline \hline  
\multicolumn{1}{||c||} {\rm  generalized picture}&
\multicolumn{3}{c||} {\rm evolution equations for}
   \\
    \cline{2-4}
   {\rm type}
 &
    $Q(t)$  & $|\psi(t)\kt$ & $|\psi(t)\kkt$
   \\
 \hline
 \hline
 \multicolumn{1}{||c||}{\rm 
 Schr\"{o}dinger, NHSP $\equiv$ DMF}  &{\rm  --}&{\rm yes}
  &{\rm yes}\\
 \hline
 \multicolumn{1}{||c||}{\rm
 Heisenberg, NHHP}  &{\rm yes} &{\rm --}
  &{\rm --}\\
 \hline
 \multicolumn{1}{||c||}{\rm
 (Dirac) interaction, NHIP} &{\rm yes} &{\rm yes}
  &{\rm yes}\\
 \hline \hline
\end{tabular}
\end{table}

\subsection{Recommended notation conventions\label{recommended-notation}}

In the most general NHIP context the use of the non-stationary,
Dyson-motivated Ansatz of eq.~(\ref{fakto}) seems to open a 
Pandora's box of thorny problems. First of all,
once we accept the fact that $\Omega=\Omega(t)$
is a map which inter-relates a triplet of Hilbert spaces,
we find that the notation is insufficient
and/or incomplete. Thus, first of all,
it does not inform us that
such a map connects the initial-reference ket
$\psi_0 \in  {\cal H}^{(\rm initial)}_{[\rm textbook]}$
and the final-reference ket
$\psi \in  {\cal H}^{\rm (user\!-\!friendly)}_{[\rm unphysical]}$, 
which may then itself be 
re-read, alternatively, as the correct physical ket
$\psi \in  {\cal H}^{(\rm final)}_{[\rm physical]}$.
Secondly, one would also like to
avoid using the
subscript $0$ in $\psi_0$ because the symbols $\psi$ and $\psi_0$
refer, in fact, to {\em the same\,} quantum state in
the NHIP representation.

Both of these inconsistencies of notation
were successfully removed in ref.~\cite{Znojil_SIGMA_2009},
in which it was shown how all of the unnecessary repetitions of 
the, otherwise necessary, explanatory comments 
may be circumvented by
using the following three simple amendments of the
standard Dirac notation for state vectors, viz., 
by using the triplet of replacements
 \be
 \psi \to |\psi\pkt\,,
 \quad \psi_0 \to  |\psi\kt\,,
 \quad \Theta\,\psi_0 \to  |\psi\kkt \ \equiv \
 \Theta\,|\psi\kt\,.
 \label{3kets}
 \ee
The defining relation between the ket state
$|\psi(t)\kt$ and the ketket state $|\psi(t)\kkt$ in
eq.~(\ref{3kets}) validates our earlier assertion 
in eq.~(\ref{nestac}) that the HP metric operator
$\Theta_{(\rm HP)}$ is stationary since,
by definition (and see table~\ref{trewe}),
neither of these states evolves in time in an HP formalism.

Continuing with the exposition of 
our recommended notation convention in the three-Hilbert-space
approach, we note that the NHIP version of our fundamental Ansatz 
of eq.~(\ref{fakto}), viz.,
 \be
 |\psi(t)\pkt = \Omega_{}(t)\,
 |\psi^{}_{}(t)\kt\,,
 \label{anza}
 \ee
in terms of the ket state $|\psi(t)\kt$, 
may equivalently be re-written as 
 \be
 |\psi(t)\pkt = \left [
 \Omega^\dagger(t)\right ]^{-1}
 |\psi^{}_{}(t)\kkt\,,
 \label{arenza}
 \ee
in terms of the ketket state $|\psi(t)\kkt$.
The unphysical nature of
the auxiliary Hilbert space is now evident
because the mean value of 
the operator representing a self-adjoint (textbook)
lower-case observable [say, $\mathfrak{q}(t)$]
becomes different from that of 
its upper-case auxiliary-space counterpart,
 \be
 \pbr \psi(t)|\mathfrak{q}(t)|\psi(t)\pkt
 = \br \psi(t) | \Omega^\dagger(t) \mathfrak{q}(t) \Omega(t) | \psi(t) \kt
 \label{reinte:a}
 \ee
 \be
 =
 \br \psi(t) | \Theta(t) Q(t)  | \psi(t) \kt
 \neq \br \psi(t) | Q(t) | \psi(t) \kt\,.
 \label{reinte:b}
 \ee
In the derivation we have employed the definition
of $Q(t)$ given in eq.~(\ref{obslambda}).
Using our notations we thus also reveal the correct physical status of
the final Hilbert space ${\cal H}^{(\rm final)}_{[\rm physical]}$,
since eqs.~(\ref{reinte:a}) and (\ref{reinte:b}) clearly
display the {\it equality\,} of the {\it measurable\,} quantities in the two respective Hilbert spaces ${\cal H}^{(\rm initial)}_{[\rm textbook]}$ 
and ${\cal H}^{(\rm final)}_{[\rm physical]}$,
 \be
 \pbr \psi(t)|\mathfrak{q}(t)|\psi(t)\pkt=\bbr \psi(t) | Q(t)
  | \psi(t) \kt\,.
  \label{remeasu}
  \ee
The required 
matrix element in the latter space [i.e., that on the right-hand side 
of eq.~(\ref{remeasu})] is precisely the one that we introduced 
earlier in the discussion in sect.~\ref{NHHP}.

We never need to leave the auxiliary Hilbert space. One can say even more:
thus, any physical quantum state can be characterized by ket
$|\psi(t)\kt$ and metric $\Theta(t)$ or,
much more economically, by the pair comprising ket $|\psi(t)\kt$ and
ketket $|\psi(t)\kkt$.
Similarly, one can say that
any physical observable can be characterized by its 
``hiddenly-Hermitian'' operator
$Q(t)$
and metric $\Theta(t)$ or,
without an explicit use of the metric, by the {pair of} 
operators
$Q(t)$
and $Q^\dagger(t)$, 
mutually connected by the metric,
 \be
 Q_{}^\dagger(t)\,\Theta_{}(t)=
 \Theta_{}(t)\,Q_{}(t)\,.
 \label{quasiH-Lambda:a}
 \ee
Equation~(\ref{quasiH-Lambda:a}), which is easily derived from 
the definitions of $\Theta(t)$ and $Q(t)$ in eqs.~(\ref{product})
and (\ref{obslambda}), respectively, is
just the so-called hidden-Hermiticity 
(alias quasi-Hermiticity) relation \cite{Scholtz-Geyer-Hahne_1992,Dieudonne_1961}
for an operator $Q(t)$ in the NHIP formalism
that belongs to a physical observable.  It is, of course,
completely equivalent to the relation 
 \be
 \Theta^{-1}(t)Q^{\dagger}(t)\Theta(t) \equiv
 Q^{\ddagger}(t)=Q(t)\,,
 \label{quasiH-Lambda:b} 
 \ee
which provides simply the obvious generalization of
the definition of $H^{\ddagger}$ 
in the flowchart of eq.~(\ref{THSstac}).
It is worth noting that the operator $\Theta$ itself is 
{\em both\,} Hermitian in the ordinary sense {\em and\,} 
quasi-Hermitian in the sense of eq.~(\ref{quasiH-Lambda:b}).  More generally, 
so is any operator that may be expressed as an arbitrary power 
series in $\Theta$ with real-valued coefficients.

As an aside here, we can now directly compare the present NHIP formalism
with the earlier NCCM description.  Thus, in the NCCM the reference 
(or model) state $\psi_0$ is assumed to be {\em stationary\,} (i.e.,
time-independent).  By comparison of eqs.~(\ref{NCCM-ev:a}) and 
(\ref{CCM-gen-sim-Xm}) with eqs.~(\ref{reinte:a}) and (\ref{reinte:b},),
we also see that the NCCM operator $\tilde{S}(t)$ is (up to a
multiplicative normalization constant) precisely equal to the
NHIP metric operator, viz., $\tilde{S}(t)=\Theta(t)/\br\psi(t)|\Theta(t)|\psi(t)\kt$,
in our newly recommended NHIP notation.

\subsection{Evolution equations
for states}

In the light of the definition, given by  eq.~(\ref{obsham}), 
of the isospectral non-Hermitian partner $H(t)$ of the textbook
Hamiltonian $\mathfrak{h}$, 
one can now claim that the upper-case operator $H(t)$ is, indeed,
quasi-Hermitian,
 \be
 H_{}^\dagger(t)\,\Theta_{}(t)=
 \Theta_{}(t)\,H_{}(t)\,,
 \label{quasiH-Ham}
 \ee
in the sense of  refs.~\cite{Scholtz-Geyer-Hahne_1992,Dieudonne_1961,Williams_1969},
and in accord with eq.~(\ref{quasiH-Lambda:a}).  
Such an observable for the 
Hamiltonian is now to be interpreted as an instantaneous
total energy. 
Its observability status reflects the observability of its
(manifestly Hermitian in the usual sense)
isospectral partner $\mathfrak{h}_{}=\mathfrak{h}_{}^\dagger$.

Several authors 
(ranging from B\'{\i}la~\cite{Bila-phdthesis_2008,Bila-arXiv_2009} 
up to his most recent followers \cite{Fring-Moussa_2016a}) 
decided to re-assign the status of Hamiltonian to another operator. 
The explanation of this
slightly surprising decision is  purely terminological.
A detailed disentanglement of the puzzle has been given 
recently \cite{Znojil_AnnPhys_2017}, which we now
briefly summarize since it is very pertinent to 
our own further developments here.
It starts from the conventional initial-space SP evolution
equation given by eq.~(\ref{jednicka}), now rewritten in our new
notation of eq.~(\ref{3kets}) as
 \be
 {\rm i}\partial_t|\psi(t)\pkt=\mathfrak{h}\, |\psi(t)\pkt\,,
   \label{speq}
 \ee
and thence from its replacement by the preconditioned NHIP alternative 
defined in
${\cal H}^{(\rm user\!-\!friendly)}_{[\rm unphysical]}$. The latter,
equivalent 
Schr\"{o}dinger-like evolution
equation for ket states $|\psi(t)\kt$,
 \be
 {\rm i}\partial_t|\psi(t)\kt=G(t)\, |\psi(t)\kt\,;
   \label{bespeq}
 \ee
 \be
 G(t) \equiv H(t) - \Xi(t)\,,
\quad
 {\Xi}_{}(t)
 \equiv {\rm i}\Omega^{-1}_{}(t)\left [\partial_t\Omega_{}(t)\right ]
  \,.
 \label{generato}
 \ee
is easily derived from eq.~(\ref{anza}). It
contains, naturally, the generator of evolution which is composed of
the energy operator $H_{}(t)$  in combination with the
so-called Coriolis operator $\Xi(t)$ \cite{Znojil_ijtp_2013},
as we observe explicitly in eq.~(\ref{generato}).
For this reason, the re-assignment of
the name of the Hamiltonian from $H_{}(t)$  to the newly constructed
operator $G(t)$, as has been suggested by several authors 
(see, e.g., 
refs.~\cite{Bila-phdthesis_2008,Bila-arXiv_2009,Fring-Moussa_2016a})
as mentioned above,
is, in our view, both misleading and unfortunate. 

In our notation any given quantum state is labeled by
{\em the same\,} Greek letter 
(say, $\psi$). 
Hence, we conclude that once we are
describing the state in terms of
a {\em pair\,} of independent vectors
$|\psi(t)\kt$ and $|\psi(t)\kkt$,
we may utilize eq.~(\ref{3kets}), which gives the relationship
between them, together with eqs.~(\ref{quasiH-Ham}), 
(\ref{bespeq}), and (\ref{generato}),
to derive the complementary, second, {\em independent\,} evolution law for 
ketket states $|{\psi}_{}(t)\kkt$,
 \be
 {\rm i}\p_t|{\psi}_{}(t)\kkt =
  G_{}^\dagger(t)\,|{\psi}_{}(t)\kkt\,.
 \label{SEtaujobe}
 \ee
In this equation,
nobody has, as yet, further re-assigned
the traditional name of the
Hamiltonian to $G^\dagger(t)$.
Thus, for our current purposes we will reserve the name 
of an ``evolution generator'' for {\em both\,} of the operators
$G(t)$ and $G^\dagger(t)$.

\subsection{Evolution equations for observables
\label{evoequ}}

Given the NHIP Coriolis operator and the initial (i.e., at time $t=0$)
values of the metric ${\Theta}_{{}}(0)$ and of an arbitrary observable operator
${Q}_{{}}(0)$, such that
 \be
 {Q}^\dagger_{{}}(0){\Theta}_{{}}(0)
 ={\Theta}_{{}}(0){Q}_{{}}(0)\,,
 \label{obco}
 \ee
we may reconstruct the full time-dependence of ${Q}_{{}}(t)$
via the following NHIP version of the Heisenberg equation,
 \be
 {\rm i}\partial_t {Q}_{{}}(t) = {Q}_{{}}(t) {\Xi}_{{}}(t)
 -{\Xi}_{{}}(t) {Q}_{{}}(t) \equiv [Q(t),\Xi(t)]
 \,.
 \label{fdeveta}
 \ee
We may also work with its conjugate version for
complement $Q^{\dagger}(t)$,
 \be
 {\rm i}\partial_t {Q}^\dagger_{{}}(t)
 = [Q^{\dagger}(t),\Xi^{\dagger}(t)]
 \,.
 \label{fdevetbe}
 \ee
In the light of
eqs.~(\ref{quasiH-Lambda:a}),
(\ref{fdeveta}) and
(\ref{fdevetbe}),
the initial condition 
of eq.~(\ref{obco}) at time $t=0$
will guarantee the observability status 
of ${Q}_{{}}(t)$
at all times, as required of course, as we now show explicitly.
In particular, if we define the operator,
 \ben
  {\cal Z}(t) \equiv Q^{\dagger}(t) \Theta(t) - \Theta(t) Q(t)\,,
 \een
which is {\em not\,} an observable, and hence does {\em not\,}
evolve as given by eq.~(\ref{fdeveta}),
we may use eqs.~(\ref{fdeveta}) and (\ref{fdevetbe}),
together with the definition of $\Omega(t)$ in eq.~(\ref{product}),
to show directly that its actual evolution equation is instead given by
 \ben
 {\rm i}\partial_t {\cal Z}(t) = {\cal Z}(t)\Xi(t) 
   - \Xi^{\dagger}(t) {\cal Z}(t)\,,
 \een
which clearly has the solution ${\cal Z}(t)=0$ for all 
times $t$ under the initial condition ${\cal Z}(0)=0$,
as given by eq.~(\ref{obco}).
Indeed, straightforward differentiation of the definition
of $Q(t)$ in eq.~(\ref{obslambda})
(and of its conjugate) with respect to time yields immediately
(the two commutator-containing) eqs.~(\ref{fdeveta}) and
(\ref{fdevetbe}). Naturally, we assume 
in so doing that the inaccessible
SP operator $\mathfrak{q}$ does not vary with time. 
We do not consider further here the
strongly anomalous case of a manifestly time-dependent 
operator, $\mathfrak{q}=\mathfrak{q}(t)$, 
except to note that it has also been discussed in
ref.~\cite{Znojil-pla_2015}.

Let us now return to eq.~(\ref{bespeq}) [skipping an analogous analysis 
of its adjoint partner in eq.~(\ref{SEtaujobe})] 
and let us notice that it contains the
operator $G(t)$
of eq.~(\ref{generato})
which is not only 
non-Hermitian but even non-quasi-Hermitian in our working Hilbert space
${\cal H}^{(\rm user\!-\!friendly)}_{[\rm unphysical]}$.
It is worth adding that it
plays not only the role of
the generator of the evolution of the state-vector kets,
as exhibited in eq.~(\ref{bespeq}),
but also the role of the
generator of the evolution of the observable instantaneous energy.
It is easy to verify the validity of the evolution equation
 \be
 {\rm i}\partial_t {H}_{{}}(t) =
 [G(t),H(t)]=[H(t),\Xi(t)]
 \,,
 \label{hadeveta}
 \ee
where the last equality, which follows simply from the
definition of $G(t)$ in eq.~(\ref{generato}), 
is thus completely compatible with eq.~(\ref{fdeveta}).
Equation~(\ref{hadeveta}) thus determines the Hamiltonian $H(t)$
from any initial value $H(0)$ such that
$H^\dagger(0)\Theta(0)=\Theta(0)H(0)$.

Even in the case of  
strictly unitary evolution
and even in the HP subcase of the general (and entirely
methodical) NHIP  scheme, as outlined above, 
both of the operators
$\Xi(t)$ and
$\Xi^\dagger(t)$
should properly be called Coriolis-force potentials 
(or, more simply, Coriolis terms) rather than Hamiltonians.
The name Hamiltonian should, for reasons thus outlined,
remains consistently restricted only to
the quasi-Hermitian observable  $H(t)$ and/or, whenever needed, 
to its conjugate operator $H^\dagger(t)$.
This is only reinforced by the 
fact that none of the other,
rejected candidates (from those introduced above) for 
possible promotion to Hamiltonian status
even need to have a real eigenvalue spectrum in general, as may
rather easily be illustrated.

Thus, for example, we may readily prove the relation,
 \ben
 {\rm i}\Omega(t)
 \left [
 \Xi^\dagger(t)-\Xi(t)
 \right ]
 \Omega^\dagger
 =
 \Omega(t)\partial_t\Omega^\dagger(t)+
 \left [\partial_t\Omega(t)\right ]\Omega^\dagger(t)
 =\partial_t
 \left [
 \Omega(t)\Omega^\dagger(t)
 \right ]\,,
 \een
simply by making use of the definition of $\Xi(t)$ from
eq.~(\ref{generato}).
This relation  implies that the (generally) non-Hermitian 
Coriolis operator $\Xi(t)$ has real eigenvalues [i.e, when it becomes
Hermitian, $\Xi^{\dagger}(t)=\Xi(t)$]
only in the special case in which the self-adjoint image $\theta(t)$
in the original space ${\cal H}^{(\rm initial)}_{[\rm textbook]}$
of the self-adjoint 
Hilbert-space metric $\Theta(t)$ in the final
space ${\cal H}^{(\rm final)}_{[\rm physical]}$, defined,
exactly as in the general case of eq.~(\ref{obslambda}), as
$$\theta(t) \equiv \Omega(t)\Theta(t)\Omega^{-1}(t)
=\Omega(t)\Omega^\dagger(t)=\theta^\dagger(t)\,,$$ 
remains time-independent.

\section{Physics behind non-stationary non-Hermitian operators \label{uctvrta}}

The initial motivation of our interest in the 
mutual relationship between the CCM approach 
to quantum many-body theory and the so-called three-Hilbert-space 
formulation of quantum theory was both formal and physical.  We noticed that 
in both of these approaches one tries to 
combine the idea of the availability of a 
straightforward,
formally friendly, approximation (or of a sequence
of approximations) of $\psi$ by a simpler $\psi_0$,
with the practical awareness of the weak points 
(i.e., typically, of a slowness of convergence) of the respective approach.
For a formal remedy to the latter weakness 
one may then immediately turn to the, by now, rather standard 
mathematical technique of the so called preconditioning 
\cite{Benzi_2002,Chen-book_2005,van-der-Vorst-book_2009,Liesen-Strakos-book_20012}, 
the essence of which lies in the use of the factorized Ansatz of eq.~(\ref{fakto}).

 \subsection{Physics of time-dependent correlations\label{ibintro}}

For both the CCM and three-Hilbert-space formalisms, 
successful examples of
the practical implementation of the
Hilbert-space-mapping approach abound. 
In just the specific context of
many-fermion quantum physics, for example, 
we may recall, e.g.,
the pioneering works 
of Dyson \cite{Dyson_1956a,Dyson_1956b},
Maleev \cite{Maleev_1957}, Coester {\em et al.} \cite{Coester_1958,Coester-Kummel_1960}, 
\v{C}\'{\i}\v{z}ek 
{\em et al.} \cite{Cizek_1966,Cizek_1969,Cizek-Paldus_1971}, 
and Janssen {\em et al.} \cite{Janssen-et-al_1971}, 
in all of which the above broad approach was adopted.
The subjects and applications of these papers covered a broad spectrum 
of applied quantum theory, including
condensed-matter physics \cite{Dyson_1956b,Bishop-Li-Campbell_2014},
nuclear physics 
\cite{Arima-Iachello_1976,Arima-Iachello_1978,Arima-Iachello_1979,Kummel_Luhr-Zab_1978}, 
and the descriptions 
of large atoms and molecules in quantum chemistry 
\cite{Monkhorst_1977,Bartlett-Musial_2007}. 

The essence of all of these techniques of making predictions concerning complicated 
multi-particle structures may be said to lie in a combination of a systematic 
mathematics behind  approximations (reflecting, typically, the calculation 
feasibility aspects) with 
an intuitive insight in the relevance of various competing phenomena 
(i.e., typically, of the various types of the  correlations). 
Thus, in a way which motivated our present paper, a transfer of the 
years-long experience from the static to non-stationary systems and 
calculations may be perceived as a true challenge. 

We came to the conclusion that a key component of such a transfer should 
be seen in a clear abstract formulation of the formalism itself. 
In this sense, we tried to propose and promote a combination of the 
NHIP language (emphasizing, in essence, the clear separation of 
the state- and operator-evolution equations)
with the CCM-related constructive efficiency. 
We believe that such a combination will pave the way towards 
an increase of efficiency of the calculations, based on the  use of  
creation and annihilation operators which is, 
simultaneously, linear (i.e., related to the construction of the bases) 
and non-linear (entering the mimicking of correlations via the 
judicious forms of the exponential mappings $\Omega$).

As we have already noted,
at various levels of approximate implementation 
of any version of the CCM a loss of manifest Hermiticity between
corresponding ket and bra states can arise
from one of its basic tenets, viz., that formally
the ground ket and bra
states are parametrized independently 
\cite{Arponen_1983,Bish-Arp-Paj_1989}. 
In practice any potential shortcomings
that this may give rise to
are far outweighed by the fact that in so doing 
one ensures the exact maintenance of the important
Hellmann-Feynman theorem \cite{Hellmann_1935,Feynman_1939} 
at all such levels of 
approximation (and see ref.~\cite{Bishop_1998} for details).
Thus, calculations
of the ground-state expectation
values of a Hermitian operator $\mathfrak{q}$ that
represents an arbitrary observable quantity
are completely compatible with 
those for the energy expectation value, in the 
sense that the former can be obtained from the
usual (perturbation-theoretical) 
Goldstone diagrams for the energy by replacing in turn
each interaction term in $\mathfrak{h}$ by the operator $\mathfrak{q}$.

\subsection{The role and interpretation of the creation and 
annihilation operators\label{create-annih-ops}}

One of the strongest motivations of 
our present goals of uncovering
a reinterpretation and translation of the 
TD-CCM techniques into the NHIP language 
(and {\it vice versa}) may be sought in 
the key merits of the CCM recipe in which one 
relies, heavily, upon the 
traditional explicit use 
of creation and annihilation operators,
as discussed in detail in sec.~\ref{CCM-fundamentals}. 
This is a decisive merit which
reflects our experience, intuition 
and pragmatic perception of the underlying phenomenology,
especially when one deals with correlations carrying
certain characteristic features of the quantum particle clustering.
Moreover, the explicit use of the concept of clusters also
leads to multiple vital simplifications of the explicit 
constructive calculations.

In the opposite direction, in the NHIP 
language one shifts attention to several related abstract
concepts, such as the 
correspondence principle and some of its less 
obvious consequences \cite{Mostafazadeh-ijgmmp_2010,Mostafazadeh-PRD_2018}. 
The conservation of probabilities,
which is guaranteed by the underlying unitarity of the temporal
evolution in our non-Hermitian pictures
is thence put in a new perspective, especially because in the
non-stationary dynamical regime
the energy clearly then ceases to be conserved.
The related operator 
will represent just the instantaneous total energy.

The NHIP framework seems 
more (perhaps, even too) general
and abstract in most application-oriented contexts.
At the same time we saw previously \cite{Bishop-Znojil_2014},
that
the very essential 
advantage gained by building bridges between the NHIP-related and CCM-related 
constructions could be viewed as
a certain optimal balance between
the abstract concepts making use of an explicit reference to the specific features 
of the systems and operators in question.

The respective implementations of the 
Ansatz of eq.~(\ref{fakto})
were largely inspired by the 
traditional Hartree-Fock methods, which were themselves
based on the choice of an
(often interaction-independent) Slater determinant $\psi_0$.
These approximations were then 
systematically upgraded
to, e.g., the most sophisticated non-Hermitian versions
of the CCM as 
described and applied to a wide variety of physical systems
in, e.g.,  refs.~\cite{Arponen_1983,Arp-Bish-Paj_1987a,Arp-Bish-Paj_1987b,Bish-Arp-Paj_1989,Bishop_1991,Bishop_1998,Coester_1958,Coester-Kummel_1960,Cizek_1966,Cizek_1969,Cizek-Paldus_1971,Bishop-Li-Campbell_2014,Kummel_Luhr-Zab_1978,Monkhorst_1977,Bartlett-Musial_2007,Bishop-Luhrmann_1978,Emrich_1981a,Emrich_1981b,Bishop-Luhrmann_1982,Kuemmel_1983,Hsue-et-al_1985,Altenbokum-Kuemmel_1985,Hasberg_Kuemmel_1986,Kaulfuss-Altenbokum_1987,Funke-et-al_1987,Arponen-Bishop_1991,Stanton-Bartlett_1993,Bishop-Kendall-Wong-Xian_1993,Arponen-Bishop_1993a,Arponen-Bishop_1993b,Funke-Kuemmel_1994,Baker-Bishop-Davidson_1996,Bish-Dav-Qu-vdW_1996,Baker-Bishop-Davidson_1997,Arponen_1997,Zeng_et-al_1998,Ligterink-et-al_1998,Ligterink-et-al_2000,Bishop-Davidson-et-al_2000,Farnell-Bishop_2004,Hagen-et-al_2014,Bishop-et-al_2017,Bishop-et-al_2019}. 
For a broad and general overview of the CCM and its applications
we may recommend the interested reader to the specific reviews 
contained in refs.~\cite{Bishop_1991,Bishop_1998,Bartlett-Musial_2007,Bishop-Kummel_1987}.

We discussed in sec.~\ref{CCM-fundamentals}} how
the NCCM and ECCM parametrizations of an arbitrary quantum many-body theory 
enable it to be
mapped {\it exactly\,} onto a classical Hamiltonian mechanics
for the many-body, classical ($c$-number) configuration-space amplitudes 
that completely and exactly describe the ket and bra ground states 
independently.  As we saw explicitly,
these mappings arise fundamentally
from an underlying one-to-one correspondence  that can be proven
to exist between the set of commutators in the original quantum many-body 
Hilbert space and a set of suitably defined generalized 
classical Poisson brackets.

It has further been shown 
\cite{Arp-Bish-Paj_1987a} how the CCM (particularly
in its ECCM form) can be interpreted 
as an {\it exact\,} generalized mean-field 
theory (i.e. beyond the Hartree-Fock level) formulation 
of the given quantum many-body problem. This
interpretation is itself closely linked with the additional
realization that the ECCM can also be construed as an 
{\it exact\,} bosonization procedure in which the ECCM states
are associated in a one-to-one fashion with a set of
generalized coherent states in some suitably defined boson
space. This ECCM bosonization procedure differs from
other such procedures in the sense that the usual motivation 
for any bosonization scheme is taken to its logical
conclusion, viz., that the resultant generalized
coherent boson fields are classical $c$-number fields with
only classical (but highly nonlinear) interactions 
between them. Being able to reinterpret the ECCM as an exact
generalized mean-field theory is then reinforced
by being able to show that, within the ECCM bosonization scheme, 
commutators of operators in the original Hilbert space 
are mapped only onto the {\it tree-level\,} pieces of the 
corresponding commutators of the
respective mapped operators in their boson image space.
The tree level of a commutator is here defined 
to be a restriction only to such contractions 
that do not result in closed loops.
The subsequent manifest exclusion of all closed-loop 
diagrams thus acts to further reinforce 
the fact that the ECCM exactly
reformulates the quantum-mechanical
many-body system that we start with 
as a classical generalized mean-field theory.

\section{Illustrative examples of the TD-CCM approaches
\label{models}}

\subsection{The Rabi model revisited}
\label{Rabi-rev}

In view of our general discussions in sects.~\ref{CCM-fundamentals} 
and \ref{NHevolu-eqns} of the fundamentals of the TD-CCM
and NHIP formalisms, respectively, and the overlaps 
that we have seen exist between them, we can now appreciate
how the use of the multi-configurational
creation operators $\{C_I^+\}$ and their destruction counterparts
$\{C_I^-\}$ in the NCCM parametrizations of eqs.~(\ref{NCCM-states})
and (\ref{NCCM-ops}) leads to a technically feasible implementation
of the NHIP concepts, via the associated use of the $c$-number cluster
correlation coefficients defined in eq.~(\ref{NCCM-ops}), as
we now again illustrate.

Thus, firstly, in sect.~\ref{Raa-sub} we outlined the basic ideas behind a
prototypical application of the stationary version of the 
NCCM to the specific case of the Rabi model, whose Hamiltonian 
is as specified in eq.~(\ref{Rabi-Hamiltonian}).  The model has also
been studied in the non-stationary regime using the time-dependent version 
of the NCCM \cite{Bishop-Emary_2001}, in which the basic coefficients
$\{s_n^{(i)},\tilde{s}_n^{(i)}; i=1,2\}$ of eqs.~(\ref{NCCM-op-S:Rabi})
and (\ref{NCCM-op-Stilde:Rabi}) now become time-dependent, and the problem
reduces in practice to solving their evolution equations (\ref{NNCCM-evol-eqs})
in the same SUB-$N$ approximation hierarchy discussed in sect.~\ref{Raa-sub}.

Perhaps the most important atomic quantity of interest associated with
the Rabi model, particularly since it is amenable to experimental
observation \cite{Rempe-Walther-Klein_1987}, is the so-called {\em atomic inversion\,}, viz.,
$\br\sigma^z\kt(t)$.  It is easy to show that in the NCCM parametrization
of sect.~\ref{Raa-sub} it may be calculated as follows,
 \begin{equation}
 \langle\sigma^z\rangle(t) = -1+2\sum_{n=1}^{\infty}\tilde{s}^{(2)}_{n}(t)\,
 s^{(2)}_{n}(t)\,. 
 \label{AI} 
 \end{equation}
Similarly, the most important field observable is the photon number, whose
associated operator is $\mathfrak{n}=b^{\dagger}b$. Once again, it is easily shown
that its time-dependent expectation value has the simple NCCM parametrization,
 \begin{equation}
 \langle\mathfrak{n}\rangle(t) = \sum_{n=1}^{\infty}n\tilde{s}^{(1)}_{n}(t)\,
 s^{(1)}_{n}(t) + \sum_{n=2}^{\infty}(n-1)\tilde{s}^{(2)}_{n}(t)\,
 s^{(2)}_{n}(t)\,. 
 \label{photon-number-av} 
 \end{equation}
As was done in ref.~\cite{Bishop-Emary_2001}, 
an obvious choice of initial condition at time $t=t_0$ to illustrate the
efficacy of the method is to start the system in
the reference state itself, i.e., the state $|0\kt|\!\downarrow\kt$, 
which comprises an empty field mode and 
an unexcited atom, for which $s_n^{(i)}(t_0)=0=\tilde{s}_n^{(i)}(t_0)$ for $i=1,2$.
 
As observables, both $\br\sigma^z\kt(t)$ and 
$\br\mathfrak{n}\kt(t)$ should always be
real. However, the SUB-$N$ truncations of the NCCM cluster operators
inevitably leads to the exact Hermiticity of the corresponding bra
and ket states being broken, which means that observables
are not constrained to be manifestly real at any such level
of approximation.  Calculations show (and see ref.~\cite{Bishop-Emary_2001})
that for small values of the coupling $g$ the imgaginary parts of the
observables are extremely small, and decrease as the truncation
index $N$ is increased. Thus, for relatively small 
couplings the evolutions of the calculated values of $\br\sigma^z\kt(t)$ and 
$\br\mathfrak{n}\kt(t)$ clearly show \cite{Bishop-Emary_2001}
a (quasi-)periodic exchange of energy between the atom and the field.
Nevertheless, as $g$ is increased, the 
restriction on the maximum SUB-$N$ level that can be 
attained in practice for higher couplings, ensures that
any spurious complex parts cannot be entirely eliminated.
Thus, although the source of the errors is well known, 
a mathematically robust and well-founded
means to suppress, and eventually eliminate, them remains unknown. One of our
hopes now is that a resolutions of the error-control puzzle might be provided
by establishing a deeper correspondence between the
NHIP-related abstract ``bookkeeping'' of the relevant operator
evolution equations, and the various pragmatic CCM
solution techniques. A best possible outcome would thereby be to
improve upon the standard CCM SUB-$N$ truncation hierarchy.  We
return to this point below after discussing our second
illustrative example of an application of the TD-CCM formalism.

Before doing so, however, we remind the reader that the above simple 
choice of initial conditions 
was made purely for ease.  Other choices are certainly possible, and 
for an arbitrary physical state as the starting state, we can always 
first calculate its NCCM representation to yield
its corresponding coefficients $\{s_n^{(i)},\tilde{s}_n^{(i)}; i=1,2\}$,
which may then be taken as the initial conditions at $t=t_0$.

\subsection{Condensed Bose fluid}
\label{Bose-cond}

We now turn to our second, much more ambitious, example of an application of the 
TD-CCM, viz., to describe the zero-temperature hydrodynamics of a macroscopic
condensed Bose fluid, this time via the ECCM \cite{Arp-Bish-Paj-Rob_1988}.
A convenient choice of reference state for this model is now $|\psi_0\kt=
|0\kt$, the bare vacuum. Accordingly, we are thus working in
a number-nonconserving formulation, as introduced by Bogolubov 
\cite{Bogolubov_1947}. Particle-number conservation is then imposed 
by working with the grand canonical Hamiltonian $\mathfrak{k}_{(0)}
\equiv \mathfrak{h}_{(0)}-\mu\mathfrak{n}$ rather than
with the Hamiltonian, $\mathfrak{h}_{(0)}$, 
where $\mu$ is the chemical potential and $\mathfrak{n}$ is the particle
number operator. We consider here a system of $N$ identical bosons, each of mass
$m$ and interacting via pairwise potentials, such that 
(in first quantization),
\be
 \mathfrak{h}_{(0)}=-\frac{1}{2m}\sum_{j=1}^{N}\nabla_j^2
 +\sum_{i=1}^N\sum_{j<i}\mathfrak{v}({\rm x}_i - {\rm x}_j)
 \label{Bose-ham-first-quant}\,.
\ee

We parametrize the many-body configuration space in terms
of real-space coordinates, such that the single-boson creation and destruction
operators, $b^{\dagger}({\rm x})$ and $b({\rm x})$, respectively, 
which act to create or annihilate a particle at the three-space
point ${\rm x}\equiv (x_1,x_2,x_3)$, obey the usual bosonic commutation 
relations, $[b({\rm x}),b^{\dagger}({\rm y})]=\delta^{(3)}({\rm x}-{\rm y})\mathbbm{1}$.
In the original Hilbert space with fixed particle number $N$,
eq.~(\ref{identity-resolution}) now takes the specific form,
 \be
 \mathbbm{1}=\sum_{n=0}^{\infty}
 \frac{1}{n!}\int {\rm dx}_1\cdots\int {\rm dx}_n\,
 b^{\dagger}({\rm x}_1)\cdots b^{\dagger}({\rm x}_n)
 |0\kt\br0| b({\rm x}_n)\cdots b({\rm x}_1)\,.
 \label{identity-bosons}
 \ee
Correspondingly, the generic ECCM amplitudes of eqs.~(\ref{ECCM-Sigma-tilde-op})
and (\ref{ECCM-Sigma-op}) now also take the specific forms,
\begin{subequations}
 \begin{align}
  \Sigma(t) &= \sum_{n=1}^{\infty}\frac{1}{\sqrt{n!}}
  \int {\rm dx}_1\cdots\int {\rm dx}_n\,
  \sigma_{n}({\rm x}_1 ,\cdots, {\rm x}_{n};t)\,
  b^{\dagger}({\rm x}_1)\cdots b^{\dagger}({\rm x}_n)
  \,,
  \label{Sigma:Bose-fluid}\\
  \tilde{\Sigma}(t)&= \sum_{n=1}^{\infty}\frac{1}{\sqrt{n!}}
  \int {\rm dx}_1\cdots \int{\rm dx}_n\,
  \tilde{\sigma}_n({\rm x}_1 ,\cdots,{\rm x}_{n};t)\,
  b({\rm x}_n)\cdots b({\rm x}_1)
  \,.
  \label{Sigma-tilde:Bose-fluid}
 \end{align}
\end{subequations}

We can now see immediately that the ECCM formalism of sect.~\ref{ECCM-fund}
provides a framework in which to consider the general case of a non-stationary
and spationally non-uniform condensate,
$\br b({\rm x})\kt(t)=\br\psi_0|{\rm e}^{\tilde{\Sigma}(t)}{\rm e}^{-S(t)}
b({\rm x}){\rm e}^{S(t)}|\psi_0\kt$.  Indeed, by making use of eqs.~(\ref{nestsum:a}) and
(\ref{nestsum:b}) and the fact that $S(t)$ has a corresponding expansion to that for
$\Sigma(t)$ in eq.~(\ref{Sigma:Bose-fluid}), we see that 
${\rm e}^{-S(t)}b({\rm x}){\rm e}^{S(t)}=b({\rm x})+[b({\rm x}),S(t)]$.  Finally,
use of the relation $b({\rm x})|\psi_0\kt=0$ then shows that
\be
 \br b({\rm x})\kt(t)=\br\psi_0|{\rm e}^{\tilde{\Sigma}(t)}b({\rm x})S(t)|\psi_0\kt
 =\br\psi_0|b({\rm x}){\rm e}^{\tilde{\Sigma}(t)}S(t)|\psi_0\kt
 =\sigma_{1}({\rm x};t)\,,
 \label{cond:sigma1}
\ee
where the last equality follows from the generic definition of eq.~(\ref{ECCM-Sigma-coeffts}).
Thus the condensate wave function $\br b({\rm x})\kt(t)$ is precisely given by
the one-body ECCM amplitude $\sigma_{1}({\rm x};t)$.  Thus, it is no surprise
to learn that at the lowest SUB-$1$ (i.e., one-body 
mean-field) level of truncation of the general 
SUB-$N$ hierarchy (in which all ECCM
amplitudes $\sigma_n$ and $\tilde{\sigma}_n$ with $n>N$ are set to zero) our formalism 
simply reduces to the well-known Gross-Pitaevskii\cite{Gross_1961,Pitaevskii_1961}
description of the condensate wave function (or one-body order parameter). Hence, our
usual SUB-$N$ approximation hierarchy can now be used to improve systematically 
upon this lowest-order one-body mean-field description.

In order to take full advantage of the local $U(1)$ gauge symmetry that
follows from particle-number conservation it is now very convenient 
to couple the system to external $U(1)$ {\em gauge fields\,} $\phi({\rm x},t)$
and ${\rm A}({\rm x},t)$ (viz., the scalar and vector potentials, respectively),
and hence we now study the new grand canonical Hamiltonian $\mathfrak{k}$, where
(in first quantization),
\begin{subequations}
\be
 \mathfrak{k}\equiv\sum_{j=1}^{N}\frac{1}{2m}\big[-{\rm i}\nabla_j-{\rm A}({\rm x}_j,t) \big]^2
 +\sum_{j=1}^{N}\big[\phi({\rm x}_j,t)-\mu\big]
 +\sum_{i=1}^N\sum_{j<i}\mathfrak{v}({\rm x}_i - {\rm x}_j)\,.
 \label{Bose-K-first-quant}
\ee
Equation (\ref{Bose-K-first-quant}) may be equivalently rewritten in the second-quantized form,
\begin{multline}
 \mathfrak{k}=\frac{1}{2m}\int {\rm dx}\big[\nabla_{x}b^{\dag}({\rm x})\big]\!\cdot\!
 \big[\nabla_{x}b({\rm x})\big]
 +\frac{\rm i}{2m}\int{\rm dx}\,{\rm A}({\rm x},t)\!\cdot\!\Big\{b^{\dag}({\rm x})
 \big[\nabla_{x}b({\rm x})\big]-\big[\nabla_{x}b^{\dag}({\rm x})\big]b({\rm x})
 \Big\} \\
 +\int {\rm dx}\Big[\phi({\rm x},t)-\mu+\frac{1}{2m}{\rm A}^{2}({\rm x},t)\Big]
 b^{\dag}({\rm x})b({\rm x})+
 \frac{1}{2}\int {\rm dx}\!\int {\rm dy}\,\mathfrak{v}({\rm x}-{\rm y})
 b^{\dag}({\rm x})b^{\dag}({\rm y})b({\rm y})b({\rm x})\,.
 \label{Bose-K-second-quant}
\end{multline}
\end{subequations}

Using this grand canonical Hamiltonian $\mathfrak{k}$ as the time-evolution
operator, one may now explicitly evaluate \cite{Arp-Bish-Paj-Rob_1988} the equations of motion 
(\ref{ECCM-evol-eqs}) (with $\mathbb{H}\to\mathbb{K}$) for the
ECCM amplitudes $\{\sigma_{n},\tilde{\sigma}_{n}\}$.  The whole treatment
may now be developed in a completely gauge-invariant fashion so as
to provide a complete (zero-temperature) hydrodynmical description. In particular, 
one may study the evolution of the expectation values of the  
off-diagonal one- and two-body density operators.  One can thereby
derive in a wholly gauge-invariant form the exact balance equations 
(i.e., the local conservation laws) for the 
hydrodynamically relevant variables (viz., 
the particle-number density, current
density, and energy density, all considered as functions
of spatial position and time), and show how 
these are in turn related to (and derivable from) the cumulant
expansion of the one-body density matrix.  It has also been shown
\cite{Arp-Bish-Paj-Rob_1988} that each of the balance equations is 
exactly obeyed at every SUB-$N$ level of truncation.

We note that the $U(1)$ gauge invariance that we have been at pains 
to incorporate exactly, can, of course, itself be
interpreted as the (differential or) local Galilean invariance.  An
immediate consequence is that the description correctly separates
the average translational (hydrodynamical) kinetic energy from the
average kinetic energy in the local rest frame, which has otherwise
proven to be very difficult to achieve in a fully microscopic treatment.
One can clearly reinterpret this exact feature of the time-dependent
ECCM treatment of a condensed Bose fluid with the generic proper treatment
within the NHIP of the analogous Coriolis terms $\Xi(t)$ of 
eq.~(\ref{generato}) in the evolution generator operator $G(t)$. 
Despite being beyond the scope of the present paper, it would be 
interesting in this light to pursue this relationship further, since
one might expect it to extend our intuitive insight into both methods,
as well as to lead possibly to improved computational (i.e., approximation) schemes.

Before closing this discussion, however, we note that the temporal
evolution of the system has been described by a trajectory in the ECCM 
symplectic phase space spanned by the cluster coefficients
$\{\sigma_n,\tilde{\sigma}_n\}$. The modern viewpoint of statistical mechanics
\cite{Abraham-Marsden-book_1978}, however, is that a proper
qualitative description of the system is obtained, rather, from the entire
{\em phase portrait\,}, which is just the totality of all allowed trajectories.
From this stance, it is clear that the above single-trajectory ECCM 
approach might now usefully be extended by focusing more on
the geometric properties of the ECCM phase space, as have been discussed
in some detail in ref.~\cite{Arponen-Bishop_1993b}.  By contrast, we 
have so far focused more on the algebraic structure of the ECCM phase space
(and see ref.~\cite{Arponen-Bishop_1993a} for further details),
and in so doing we have been led inexorably to the SUB-$N$ hierarchy 
of approximations.  As we have seen in sec.~\ref{Raa-sub}, for example,
this scheme sometimes has computational limitations, which stem 
essentially  from the fact that, although the ECCM has left its ultimate
origins in perturbation theory far behind, the SUB-$N$ scheme itself 
does still seem to have some links with perturbation theory.  The hope has 
been expressed \cite{Arp-Bish-Paj-Rob_1988} that one might be able to
exploit the geometric structure of the ECCM phase space 
to derive wholly new (and potentially more robust and more powerful)
approximation schemes based wholly on its geometric properties.  Again,
we note that, although such an investigation lies beyond the scope of 
the present paper, any progress in this area will certainly find
immediate impact in the NHIP formalism too.

\section{Outlook \label{ctvrta}}

In the arena of realistic and predictive CCM calculations,
most attention is usually paid to
the study of molecules and/or other quantum many-body systems
in their ground state.
Once our interest shifts to the
excited states the strategy has to be modified,
as we alluded to in sec.~\ref{druha}. Within the S-CCM
such a modification was first proposed by Emrich \cite{Emrich_1981a,Emrich_1981b} 
within the context of the 
time-independent (i.e., stationary) Schr\"{o}dinger equation. 
This was done by a suitable modification of the ground-state 
parametrization, which did not involve a change of reference state.  
A later alternative TD-CCM 
approach to excited states was advocated by Arponen 
and his co-workers \cite{Arponen_1983,Arp-Bish-Paj_1987b} within 
the context of the time-dependent Schr\"{o}dinger equation, wherein 
it was shown how excited states could be obtained as the normal modes
of a suitably defined dynamical Hamiltonian matrix obtained 
within linear response theory, in direct analogy to 
the usual procedure in classical mechanics.
The complete equivalence of the two approaches has been formally
demonstrated \cite{Arp-Bish-Paj_1987b,Bish-Arp-Paj_1989,Bishop_1998}. 
In effect, what we have done in the present paper is to build 
upon, and extend, these results.

 \subsection{Reference-dependence versus reference-independence}
 
In the broader context of
constructive quantum many-body theory
the key purpose of our present paper
may be seen in the description of the close
relationship
between the reference-dependent
and reference-independent forms of the Ansatz 
of eq.\ (\ref{fakto}). In this sense our
present study of parallels
between the alternative NHIP and TD-CCM
approaches to non-stationary dynamics
was preceded by our earlier paper \cite{Bishop-Znojil_2014}
in which the methodical framework
was
perceivably simplified and reduced to the less sophisticated
search for parallels between
the stationary DMF and S-CCM methods.

As a concise summary of the message provided 
by ref.~\cite{Bishop-Znojil_2014}
it can be said that the reconsideration of the S-CCM method
from the reference-independent DMF perspective may be
well illustrated in diagrammatic form by 
the flowchart in our present eq.~(\ref{THSstac}). 
Therein one assumes an input knowledge of the Hamiltonian operator
$\mathfrak{h}$ and of the Dyson mapping $\Omega$.
In this setting the main weakness of the
reference-independent DMF recipe lies in the
absence of any hint of how we should 
choose the operator map $\Omega$.
Hence, the stationary DMF approach may, somewhat
crudely, but simply, be characterized
as guesswork, or as an intuition-based
implementation of a purely trial-and-error-type strategy.

This does not mean that the DMF constructions were not successful
in practice.
In fact, the opposite is true \cite{Scholtz-Geyer-Hahne_1992}.
Still, the S-CCM techniques are
more systematic
because, given a Hamiltonian operator
$\mathfrak{h}$, the S-CCM specification of the
necessary Dyson operator $\Omega=\exp S$ is not at 
all arbitrary, with the rationale for the 
exponentiated form as we have outlined above.
Hence, the S-CCM techniques are both``algorithmic'' and
extremely intuitive.
A related major 
advantage of the S-CCM lies in the use of the 
very concrete truncated-series expansions of 
the Dysonian exponent operator $S$ in 
an operator
basis formed by certain multi-configurational creation 
operators, as alluded to above.
This opens the dual possibilities of multiple technical 
simplifications and extremely powerful approximation
hierarchies that are well-defined and physically motivated, both of 
which are {\em precisely due to\,}
the intrinsic reference-dependence of the S-CCM construction.

In the present paper we have pointed out 
that a similar balance between the 
merits and shortcomings must also necessarily exist 
(and should prove useful) when one 
replaces the S-CCM\,$\leftrightarrow$\,DMF stationary correspondence
by its TD-CCM\,$\leftrightarrow$\,NHIP time-dependent generalization.
{\it A priori\,} one may expect that in the 
generalized setting, {\em any\,} technical 
simplification (obtained from the mutual enrichment)
could play a more decisive role, first of all, because of the
enormous overall increase in mathematical complexity  
that emerges after
the respective replacements S-CCM\,$\to$\,TD-CCM and/or DMF\,$\to$\,NHIP.

Particular attention has to be paid to a deeper theoretical
role played by the transfer of mathematical know-how from the 
DMF and NHIP formalisms to their respective S-CCM and TD-CCM counterparts.
In this manner the brute-force S-CCM and TD-CCM constructions 
appear
open to further amendments.  A typical example is
the possible replacement of a certain {\it ad hoc}, auxiliary 
S-CCM or TD-CCM operator admitted by the principles of bivariationality, 
and hence needed to
characterize the bra states {\it independently\,} from the ket states 
(viz., the operator $\tilde{S}$ in the NCCM or 
$\tilde{\Sigma}$ in the ECCM)
that is thus required for the evaluation of 
the expectation values of an arbitrary 
physically observable operator, by  
alternative forms that maintain 
the canonical CCM symplectic structure
discussed in sec.~\ref{CCM-fundamentals}, which
are both inspired and restricted 
by the implicit mathematical merits 
(e.g., the Hermiticity reinstallation property) of its
metric-operator-provided reinterpretation.

 \subsection{Interaction-picture context}

Let us recall the well-known method of 
Seidewitz \cite{Seidewitz_2017} for avoiding 
the no-go consequences of Haag's
theorem in quantum field theory (QFT), which states that,
under the {\it usual\,} assumptions made in QFT,
any field that is unitarily equivalent to a free field 
must itself be a free field. In particular, 
Seidewitz shows how Haag’s theorem 
can be avoided when QFT is formulated using
an additional invariant {\it path\,} parameter, as well as the 
usual four-position parameters. His method 
relies on the removal of the spectral
condition, essentially by replacing the usual on-mass-shell
operator of relativistic energy and momentum by its off-shell
generalization $\hat{P}$. 
Importantly, this is 
accompanied by the preservation of
the traditional IP approach to the Dyson
perturbation expansions of
scattering matrices, thereby providing a fully 
consistent basis for performing the usual practical QFT
calculations. 

For the purposes of our present discussion,
what is important in the above Seidewitz 
construction \cite{Seidewitz_2017} is that, in order
to bypass the limitations imposed by Haag's theorem
in the standard formulations of QFT,
the frame-dependent zeroth
component of $\hat{P}$, which plays the role of 
the energy operator in conventional QFT,
necessarily becomes different from the 
newly introduced free relativistic Hamiltonian.
From a rather abstract point of view the latter idea is 
reminiscent of and, indeed, completely paralleled
by, our present introduction of the 
distinctions between several
alternative candidates to play the role of  
a non-stationary non-Hermitian Hamiltonian. 

As we have seen, only one of these candidates 
(viz., the IP operator $H(t)$) is 
hiddenly Hermitian (i.e., observable).
The other ones [viz., the IP operators $\Xi(t)$ and
$G(t)$], together with their conjugate forms
are not. In general, the differences $[G^\dagger(t)-G(t)]$ and
[$\Xi^\dagger(t)-\Xi(t)]$ do not vanish. 
In turn, this opens up a wide space
for multiple unconventional dynamical scenarios
that provide an interplay and mutual cancellations 
between the non-unitarily evolving components 
in states $\psi(t)$
and in observables $Q(t)$, respectively. Along these lines, the ultimate 
(hidden) unitarity of the whole picture is achieved.

We described the corresponding non-Hermitian
generalizations of evolution equations, 
thereby offering a firm ground for
the transfer (NHIP\,$\to$\,TD-CCM)
of the {\em operator\,} NHIP formalism into its
{\em variational\,} TD-CCM parallels.
In the present stage of development of the theory we already
understand why and how operators $G_{}(t)$ 
and $G^\dagger(t)$ control the
time-evolution of states,  as well as why and when 
their Coriolis-operator
partners ${\Xi}_{}(t)$ and ${\Xi}^\dagger_{}(t)$ 
play the role of the generators of
time-evolution of all of the {IP}-represented observables.
As two specific examples of where the opposite transfer
(TD-CCM\,$\to$\,NHIP) of experience from the TD-CCM
to its NHIP counterpart might lead to further insights we recall
the specific applications of the 
former to the Rabi Hamiltonian and the 
condensed Bose fluid, as discussed in sect.~\ref{models}.

In the context of much recent research in 
the area that has been aimed mainly at the SP
formalism, we have also clarified here the deep changes in the
role of the so-called Hilbert space metric in the generic 
situations in which it is allowed to vary with time,
$\Theta=\Theta_{}(t)$. Furthermore, having
shown here how we can, completely equivalently, 
work instead solely with
the {\em pairs\,} $\{|\psi_{}(t)\kt,|\psi_{}(t)\kkt\}$ 
of the single-state-representing kets and ketkets, 
we also thereby demonstrated how 
the explicit need for the operator $\Theta_{}(t)$ itself 
has certainly been weakened, if not altogether eliminated. 
We believe that a transfer of this experience 
to the variational context 
is potentially extremely productive. 

\section{Summary\label{summary}}

The formulation of the relationships illustrated 
in eq.~(\ref{THSecht})
was motivated, first of all, by the many phenomenological
successes of the conventional stationary special
case of eq.~(\ref{THSstac}). In the present paper, 
the emphasis was aimed rather at an innovative 
reinterpretation of the existing
non-stationary (TD-CCM) extensions of the
variational CCM techniques.

The underlying mathematical and technical details
required to transfer the
non-stationary Dyson-inspired
formalism (viz., the NHIP approach)
were outlined. 
It is precisely in this domain
where a large part of our present contributions 
are truly original in their own right. The remaining 
originality resides in bringing into juxtaposition,
within a newly developed conceptual and notational
framework that is broad and powerful enough
to encompass both,
two very powerful methodologies that have 
hitherto been seen as quite separate.  We have been
at some pains to draw parallels that might
henceforth be exploited to advance both formalisms 
and their subsequently enhanced arenas of applications. 
We have described in some detail how
the key source of the mathematical inspiration 
for our study
lies in the unconventional {\em non-Hermiticity\,} 
of Hamiltonians in
the formalisms. This gave birth earlier 
\cite{Bishop-Znojil_2014} to the
explicit, and rather fruitful, 
description of manifold conceptual 
parallels between two otherwise seemingly disparate
theoretical constructs.  These comprise,
on one hand, the very successful
stationary S-CCM formulations 
of quantum many-body theory and, on the other, their 
(at least, in principle) somewhat more ambitious, 
quasi-Hermitian \cite{Dyson_1956a,Dyson_1956b,Scholtz-Geyer-Hahne_1992,Dieudonne_1961} 
(otherwise known as pseudo-Hermitian \cite{Mostafazadeh-ijgmmp_2010} 
or ${\cal PT}$-symmetric
\cite{Bender-rpp_2007}) {\em stationary\,} analogues,
which, by now, have themselves also become
rather widely used in a variety of
applications of quantum theory.

A  non-stationary, TD-CCM version of the CCM theory
has been considered here, therefore. This was done because we believe
that, in contrast to the rather universal
Dyson-inspired quasi-Hermiticity techniques, a characteristic and
specific merit of the less universal CCM theory may be seen in
the much larger number of
very accurate calculations that it has found 
in such a wide range of applications in many diverse
subfields of quantum physics and chemistry.

Several conclusions from 
our study also appear to provide a
perceptibly deeper insight into abstract quantum theory itself.
In particular, we have shown that one can easily remove the
representation-framework restrictions as accepted both in
ref.~\cite{Znojil_SIGMA_2009} (wherein only  
wave functions evolved in time, 
i.e., in the non-Hermitian Schr\"{o}dinger picture) and
in ref.~\cite{Znojil-pla_2015} (where the transition to
the non-Hermitian Heisenberg picture, in which wave functions
remain constant in time, was analyzed). 
In other words, our present version of the
NHIP formalism (in which both wave functions and operators of
observables cease to remain constant in time) may be briefly
characterized as an immediate non-Hermitian generalization of the
interaction picture (alias the traditional Dirac picture) as
described in most standard quantum mechanics textbooks (and see, e.g.,
refs.~\cite{Constantinescu-book_1971,Messiah-book_2014,Sakurai-book_2017}), 
and as now almost universally used in quantum field theory calculations
(and see, e.g., refs.~\cite{Bjorken-Drell-book_1965,Negele-Orland-book_1998,Tomonaga_1946,Schwinger_1948,Dyson_1949}). 

One of the deep unifying features between the quasi-Hermiticity
techniques on the one hand and the CCM on the other is their
ability to be formulated in terms of a bivariational principle.
For the general case of quasi-Hermitian operators in
quantum mechanics, Scholtz {\it et al.} 
\cite{Scholtz-Geyer-Hahne_1992} showed in particular
how the introduction of the metric operator was especially
useful for the implementation of a variational principle, 
which could itself then be used for (approximate) calculational
purposes.  Exactly the same bivariational formulation of
both S-CCM and TD-CCM approaches
\cite{Arponen_1983,Bish-Arp-Paj_1989} 
has led to the extremely accurate descriptions
of a wide variety of strongly-interacting quantum 
many-body systems.

Another common feature of the quasi-Hermiticity and CCM
(particularly the ECCM) formalisms is their deep relationship
to exact bosonization mappings.  In the case of the ECCM
this has even resulted in its ultimate realization as
an exact {\it classicization\,} \cite{Arp-Bish-Paj_1987a}, as
we discussed  in some detail in sect.~\ref{create-annih-ops}.  The 
ECCM itself, as we also discussed, introduces {\it two\,}
independent preconditioning operators $\Omega$ of the
form of eq.~(\ref{expS}).  We now  fully expect that
this important feature of the ECCM might also find
applications into a further uesful, parallel extension of the 
three-Hilbert-space (alias the NHIP) formalism, which 
mirrors the NCCM\,$\to$\,ECCM extension.  However, such 
a discussion takes us far beyond the aims
of the present paper.

%
\bibliographystyle{spphys}       
\bibliography{refs}   

\end{document}